\IfFileExists{revtex4-2.cls}{%
  \documentclass[10pt,aps,prd,twocolumn,superscriptaddress,nofootinbib,floatfix]{revtex4-2}%
  \newif\ifrevtex\revtextrue
}{%
  \documentclass[twocolumn,10pt]{article}%
  \newif\ifrevtex\revtexfalse
}
\usepackage[utf8]{inputenc}
\IfFileExists{lmodern.sty}{\usepackage[T1]{fontenc}\usepackage{lmodern}}{}
\usepackage{amsmath,amssymb,bm}
\usepackage{graphicx}
\usepackage{booktabs}
\usepackage{xcolor}
\usepackage{hyperref}
\hypersetup{colorlinks=true,linkcolor=blue!50!black,citecolor=green!35!black,urlcolor=blue!60!black}

\ifrevtex\else
  \usepackage[letterpaper,margin=0.75in,top=0.9in]{geometry}
  \usepackage{amsthm}
  \numberwithin{equation}{section}
  \makeatletter
  \newcount\aff@count \aff@count=0
  \def\aff@list{}
  \def\auth@list{}
  \newcommand{\affiliation}[1]{\advance\aff@count by 1
    \expandafter\g@addto@macro\expandafter\aff@list\expandafter{\expandafter\item\expandafter[\the\aff@count]#1}}
  \newcommand{\email}[1]{\g@addto@macro\auth@list{\footnote{#1}}}
  \renewcommand{\author}[1]{\g@addto@macro\auth@list{#1}}
  \newsavebox{\abs@box}
  \renewenvironment{abstract}{\global\setbox\abs@box\vbox\bgroup\hsize=0.85\textwidth\small\noindent}{\egroup}
  \renewcommand{\maketitle}{\twocolumn[{\begin{center}
      {\Large\bfseries\@title\par}\vspace{1.2ex}
      {\normalsize\auth@list\par}\vspace{0.6ex}
      {\small\itshape\begin{list}{}{\leftmargin=1em\itemsep=0pt\topsep=0pt}\aff@list\end{list}\par}\vspace{0.6ex}
      {\small(\@date)\par}\vspace{1.5ex}
      \box\abs@box\end{center}\vspace{2ex}}]}
  \newenvironment{acknowledgments}{\section*{Acknowledgments}}{}
  \newenvironment{widetext}{}{}
  \newenvironment{ruledtabular}{\par\noindent\hrule height 0.6pt\vskip 2pt\centering}{\par\vskip 2pt\hrule height 0.6pt\par}
  \makeatother
\fi

\newcommand{\pw}[3]{{}^{#1}\mathrm{#2}_{#3}}
\newcommand{\NN}{N\!N}
\newcommand{\ND}{N\!\Delta}
\newcommand{\DD}{\Delta\Delta}
\newcommand{\Nc}{N_c}
\newcommand{\MN}{M_N}
\newcommand{\MD}{M_\Delta}
\newcommand{\bone}{\mathbf{1}}
\newcommand{\Mop}{\mathcal{M}}
\newcommand{\Ctr}{C_{\mathrm{tr}}}
\newcommand{\bh}{\hat\beta}
\newcommand{\dd}{\mathrm{d}}
\newcommand{\Tp}{\mathsf{T}}

\providecommand{\doi}{}\renewcommand{\doi}[1]{\href{https://doi.org/#1}{doi:\nolinkurl{#1}}}

\newcommand{\dM}{\delta_M}
\newcommand{\LH}{\Lambda_H}
\newcommand{\LL}{\Lambda_L}
\newcommand{\kDD}{\kappa_{\DD}}
\newcommand{\kND}{\kappa_{\ND}}
\newcommand{\Vh}{\hat V}
\newcommand{\kh}{\hat k}
\newcommand{\kaph}{\hat\kappa}
\newcommand{\Dh}{\hat{\mathcal D}}
\newcommand{\Eu}{\mathcal E}
\newcommand{\Kh}{\hat K}
\newcommand{\Jop}{\mathcal{J}}
\newcommand{\Iop}{\mathcal{I}}
\newcommand{\Xop}{\mathcal{X}}
\newcommand{\Sop}{\mathcal{S}}
\newcommand{\Bop}{\mathcal{B}}
\newcommand{\Pop}{\mathcal{P}}

\begin{document}

\title{Renormalization-group flow of nucleon--$\Delta$ contact interactions with an explicit $\Delta$: decoupling and the scale dependence of large-$N_c$ relations}
\author{Dominador F. Vaso, Jr.}
\email{dominadorjrvaso@gmail.com}
\affiliation{Department of Physics, Mindanao State University - Iligan Institute of Technology, Iligan City, 9200, Philippines}
\affiliation{Khon Kaen Particle Physics and Cosmology Theory Group (KKPaCT), Department of Physics, Faculty of Science, Khon Kaen University, Khon Kaen 40002, Thailand}

\author{Daris Samart}
\email{darisa@kku.ac.th}
\affiliation{Khon Kaen Particle Physics and Cosmology Theory Group (KKPaCT), Department of Physics, Faculty of Science, Khon Kaen University, Khon Kaen 40002, Thailand}

\author{Jingle B. Magallanes}
\email{jingle.magallanes@g.msuiit.edu.ph}
\affiliation{Department of Physics, Mindanao State University - Iligan Institute of Technology, Iligan City, 9200, Philippines}

\date{\today}

\begin{abstract}
We study the leading-order (LO) contact interactions among nucleons and $\Delta$ isobars under a Wilsonian renormalization-group flow with a sharp momentum cutoff, keeping the $\Delta$ as an explicit coupled channel, and determine which large-$N_c$ relations among the couplings are invariant under the flow. The two-channel flow with the $\Delta\Delta$ threshold inside the cutoff range is solved exactly, and the closed channels decouple with relative corrections of order $\Lambda^2/\kappa_{\Delta\Delta}^2$, where $\kappa_{\Delta\Delta}=\sqrt{2M_\Delta(M_\Delta-M_N)}$ is the closed-channel momentum. Applying the $1/N_c$ expansion of the Hartree Hamiltonian to the $N$--$\Delta$ system, we find that a generalized time reversal removes all corrections of relative order $1/N_c$, that the LO fixes eleven of the 18 S-wave couplings in terms of seven, and that the exact-SU(4) two-parameter form used in the literature differs from the LO by an operator of relative order $1/N_c^2$ with matrix elements of order one in the $\Delta$ channels. The eight spin--isospin flip relations are exact invariants of the flow, also when the central part of one-pion exchange is iterated in the S waves. The eigenchannel relations of the $NN$--$\Delta\Delta$ sectors hold only at the scale at which they are imposed. Their running is controlled by $\kappa_{\Delta\Delta}$, which is of order $N_c^0$, and by a parameter that the $NN$ data do not fix, so that the $1/N_c$ expansion does not remove it. Large-$N_c$ relations between open and closed channels therefore constrain the potential at one scale. The flip relation between the $(S,I)=(3,0)$ and $(0,3)$ channels of $\Delta\Delta$ scattering predicts equal scattering lengths at LO, which lattice QCD at heavy pion masses can test.
\end{abstract}

\maketitle

%% ================================================================
\section{Introduction}\label{sec:intro}
%% ================================================================

Chiral effective field theory (ChEFT)~\cite{ehm2009,me2011} provides a systematic framework for explaining the nuclear forces, in which pion exchange gives the long-range part of the baryon--baryon potential and contact interactions with low-energy constants (LECs) encode the short-range part. The $\Delta(1232)$ is the lowest excitation of the nucleon, with the mass splitting $\dM\equiv\MD-\MN=293$~MeV, and it is the only baryon besides the nucleon that enters two-flavor nuclear EFT as an explicit degree of freedom. The $\Delta$ has been included in the chiral two-nucleon potential since the early work of Ref.~\cite{orvk1996}. The $\Delta$-full formulation improves the convergence of the two-nucleon potential~\cite{kem2007,piarulli2015,ekstrom2018,sk2020}, and the $\Delta$ enters through its couplings to pions and through contact interactions of the $\ND$ and $\DD$ channels. In pionless EFT~\cite{hkv2020} the $\Delta$ is usually integrated out together with the pion. In both frameworks the contact couplings of the channels that contain a $\Delta$ cannot be fixed by nucleon--nucleon data, since these channels open $293$ and $586$~MeV above the $\NN$ threshold, and it was found in Ref.~\cite{sk2020} that even large values of these couplings have a negligible effect on the $\NN$ phase shifts. Two sources of information are available about them. The first is the large-$\Nc$ limit of QCD~\cite{witten1979,djm1994,djm1995}, in which the nucleon and the $\Delta$ become degenerate members of a spin--flavor tower and the two-baryon interaction acquires the contracted SU(4) structure exploited for the $\NN$ potential in Refs.~\cite{ks1996,km1997}. The large-$\Nc$ ordering of the pionless $\NN$ contact terms agrees with the data~\cite{sss2018,rss2023}, and its extension to the $\Delta$ channels~\cite{rss2025} and to the SU(3) octet and decuplet~\cite{bvs2026} reduces the number of leading order (LO) couplings from 18 to 2 in the two-flavor S-waves. The second is resonance saturation, in which the contact couplings at a cutoff of the order of the vector-meson mass are identified with the low-momentum limit of $\rho$, $\omega$ and scalar exchange, as examined for the four-nucleon operators in Ref.~\cite{egme2002}. We are unaware of a corresponding analysis for the $\ND$ and $\DD$ contact terms.

Both sources refer to a scale, and the consequences of this scale dependence are the main subject of this work. The large-$\Nc$ relations are derived for degenerate $N$ and $\Delta$, whereas in nature $\dM$ is of the same size as the momenta of the low-energy theory and the $\DD$ threshold lies far above the $\NN$ threshold. Resonance saturation refers to a cutoff near $m_V\simeq770$~MeV, whereas pionless EFT is defined below the pion mass. A relation among contact couplings is not a scale-independent statement, because the couplings run with the cutoff. The subtraction-point dependence of the large-$\Nc$ relations among pionless $\NN$ couplings has been noted in the power-divergence-subtraction (PDS) scheme~\cite{ksw1998a,ksw1998b}, where the relations hold only in a window of subtraction points~\cite{sss2018,nssv2021,rss2023}, and the review of Ref.~\cite{rss2023} observes that a consistent treatment would require the explicit $\Delta$. It is therefore necessary to establish which large-$\Nc$ relations among the $N$--$\Delta$ contact couplings are invariant under the renormalization-group (RG) flow and which acquire a scale, and to identify what controls the scale dependence. Only the former can be imposed on the $\Delta$ channels of the $\Delta$-full theories, where these couplings are otherwise free, or confronted with lattice QCD, where the $\DD$ channels are accessible at heavy pion masses and the $\Delta$ is stable, without reference to a renormalization scale.

Throughout this work, we keep the $\Delta$ as an explicit coupled channel and follow the momentum-independent contact couplings with the Wilsonian RG flow of a sharp momentum cutoff. In this mass-dependent scheme the closed channels decouple automatically as the cutoff passes below the channel momenta $\kDD=850$~MeV and $\kND=559$~MeV, with power corrections, and the two-channel flow with the $\DD$ threshold inside the cutoff range is solved exactly. The large-$\Nc$ structure of the couplings is analyzed with the Hartree Hamiltonian~\cite{km1997,ps2013} for general $\Nc$, imposed as a boundary condition at a high cutoff $\LH$ and followed down to a low cutoff $\LL$. In the two sectors in which $\NN$ couples to $\DD$ the large-$\Nc$ relations fix the relative size of the $\NN$, $\DD$ and transition couplings, which we express through the eigenchannel angle $\theta$, the angle that parametrizes the eigenvector $(\cos\theta,\sin\theta)$ of the $2\times2$ coupling matrix in the $(\NN,\DD)$ basis. A large-$\Nc$ relation such as $(C_{\DD}-C_{\NN})/\Ctr=\sqrt5$ is a fixed value of $\theta$, and the running of $\theta$ measures directly how far the flow violates the relation.

The results are of two kinds. The spin--isospin flip relations $C^{(S,I)}=C^{(I,S)}$ are exact invariants of the flow, because the two sectors of a flip pair contain the same channels and have identical loop integrals, and this remains true when the central part of one-pion exchange is iterated in the S-waves. The eigenchannel relations hold only at the scale at which they are imposed. Their running between $\LH$ and $\LL$ is governed by the difference between the open $\NN$ loop and the closed $\DD$ loop, set by $\kDD=\sqrt{2\MD\dM}$, and by one parameter of the boundary condition that the $\NN$ data do not fix. The running does not disappear at large $\Nc$, since with $\MD\sim\Nc$ and $\dM\sim1/\Nc$~\cite{djm1994} the momentum $\kDD$ is of order $\Nc^0$, so that the two loops differ at every $\Nc$, and in the Hartree counting the running term is enhanced by $\Nc^2$ relative to the splitting of the eigen-couplings. The scale dependence of large-$\Nc$ relations between open and closed channels is thus a threshold effect set by $\kDD/\Lambda$, removed only in the formal limit $\kDD\to0$, and such relations constrain the potential at one scale, in line with the observation of Refs.~\cite{bcg2002,bc2002} that large-$\Nc$ relations constrain the potential rather than the amplitude. Some of these statements follow from kinematics alone, namely the invariance of the flip relations, and the decoupling of the closed channels, which is an instance of the theorem of Appelquist and Carazzone~\cite{ac1975}. Reference~\cite{rss2025} has shown that the $\NN$ and $\DD$ sectors decouple in the unitary limit and that one of the two SU(4) parameters is then driven to zero. The new results of this work are the complete LO S-wave contact sector of the $N$--$\Delta$ system, with ten $(S,I)$ sectors and 18 couplings, the general-$\Nc$ matrix elements of the Hartree operators including the $N\leftrightarrow\Delta$ transitions, a time-reversal argument that removes every correction of relative order $1/\Nc$, the two relations that survive at relative order $1/\Nc^2$, the identification of the exact-SU(4) two-parameter form of Refs.~\cite{rss2025,bvs2026} as the LO plus an operator of relative order $1/\Nc^2$ whose matrix elements are of order one in the $\Delta$ channels, and the exact solution of the flow with its consequences for these relations. The flip relation between the $(S,I)=(3,0)$ and $(0,3)$ channels of $\DD$ scattering, which is both basis independent and RG invariant, predicts equal scattering lengths at LO, and we state what the $1/\Nc$ expansion does and does not fix about its first correction.

The present study is restricted to the contact interaction sector, that is, to the momentum-independent couplings and their flow. We take $\LH=770$~MeV and $\LL=140$~MeV as representative of the cutoffs at which saturation and pionless EFT are formulated, but the flow between them is a contact flow, and the one-pion exchange that is dynamical in this window is not included. Its inclusion requires the Wilsonian flow with a long-range potential~\cite{bb2003,birse2006,birse2007,hky2011}, in which the S-wave one-pion exchange is perturbative in the $\pw{1}{S}{0}$ channel away from the amplitude pole~\cite{ks1999} and the tensor force makes the triplet channels nonperturbative above a few hundred MeV~\cite{ntvk2005}. The contact flow is therefore exact as a statement about the contact theory and not yet a statement about nature, and Sec.~\ref{sec:twocut} states where this limits the conclusions. The renormalization condition is cutoff independence of the full $2\times2$ matrix of threshold amplitudes of each sector, including the closed-channel elements that no $\NN$ datum constrains, which is a choice that we make explicit. A second limitation is the fine tuning of the $\NN$ scattering lengths, which prevents any boundary condition at $\LH$ from predicting $a_s$ and $a_t$ and restricts the predictions to relations among couplings. Finally, the structural result that scalar and vector exchange at order $q^0$ generate no $N\to\Delta$ transition coupling is an artifact of the $q^0$ truncation, which drops the exchanges that are leading in $1/\Nc$, and it is reported as such.

The paper is organized as follows. In Sec.~\ref{sec:contacts} we set up the contact interactions among nucleons and $\Delta$'s in the operator basis of Ref.~\cite{sk2020}, define the couplings of the S-wave $(S,I)$ sectors that flow, and enumerate these sectors. In Sec.~\ref{sec:decoupling} we solve the cutoff flow of the coupled $\NN$--$\DD$ sectors with the threshold inside it and derive the decoupling of the closed channels. In Sec.~\ref{sec:largeN} the $1/\Nc$ expansion of the Hartree Hamiltonian is carried out for the $N$--$\Delta$ system, the LO relations are derived, and their behavior under the flow is determined. In Sec.~\ref{sec:twocut} the flow between the two cutoffs is examined, the saturated boundary condition at order $q^0$ is derived and its limitations are stated, and the flip relation between the $\DD$ scattering lengths and its first correction are discussed. Section~\ref{sec:summary} contains the conclusions. The conventions and the spin--isospin matrices, the cutoff integrals, the Wilsonian form of the flow with its fixed points, the SU(4) construction, the boundary-condition family and the low-momentum reduction of one-boson exchange are given in the appendices.

%% ================================================================
\section{Contact interactions among nucleons and $\Delta$'s}\label{sec:contacts}
%% ================================================================

\subsection{Leading- and next-to-leading-order potentials}\label{sec:operators}

We start with the contact interactions among nucleons and $\Delta$ isobars in the operator basis and the notation of Ref.~\cite{sk2020}. The conventions for the kinematics, the two-baryon states and the scattering equation are collected in App.~\ref{app:conv}. The spin matrices are $\bm\sigma$ for the nucleon, $\bm\Sigma$ for the $\Delta$ and $\bm S^\dagger$ for the $N\to\Delta$ transition, with the symmetric traceless tensors $S^{ij\dagger}$, $\Sigma^{ij}$ and $\Sigma^{ijk}$ built from them. The isospin matrices $\bm\tau$ (nucleon), $\bm\Theta$ ($\Delta$) and $\bm T^\dagger$ ($N\to\Delta$ transition) have the same matrix representations as $\bm\sigma$, $\bm\Sigma$ and $\bm S^\dagger$, respectively, with $\Theta^3=\mathrm{diag}(3,1,-1,-3)$, so that $\bm\Theta=2\bm I$ on the $\Delta$ in the same way as $\bm\tau=2\bm I$ on the nucleon. Their explicit forms are given in App.~\ref{app:conv}. At LO the contact potentials are momentum independent and take the following form,
\begin{equation}
\begin{aligned}
V^{(0)}_{\mathrm{ct},\NN\NN}&=C_S+C_T\,\bm\sigma_1\!\cdot\!\bm\sigma_2,\\
V^{(0)}_{\mathrm{ct},\NN\ND}&=C^{(0)}_{2,\NN\ND}\,\bm\sigma_1\!\cdot\!\bm S_2^\dagger\ \,\bm\tau_1\!\cdot\!\bm T_2^\dagger,\\
V^{(0)}_{\mathrm{ct},\ND\ND}&=C^{(0)}_{1,\ND\ND}+C^{(0)}_{2,\ND\ND}\,\bm\sigma_1\!\cdot\!\bm\Sigma_2,\\
V^{(0)}_{\mathrm{ct},\NN\DD}&=\big[C^{(0)}_{2,\NN\DD}\,\bm S_1^\dagger\!\cdot\!\bm S_2^\dagger\\
&\qquad+C^{(0)}_{3,\NN\DD}\,S_1^{ij\dagger}S_2^{ij\dagger}\big]\,\bm T_1^\dagger\!\cdot\!\bm T_2^\dagger,\\
V^{(0)}_{\mathrm{ct},\ND\DD}&=\big[C^{(0)}_{2,\ND\DD}\,\bm S_1^\dagger\!\cdot\!\bm\Sigma_2\\
&\qquad+C^{(0)}_{3,\ND\DD}\,S_1^{ij\dagger}\Sigma_2^{ij}\big]\,\bm T_1^\dagger\!\cdot\!\bm\Theta_2,\\
V^{(0)}_{\mathrm{ct},\DD\DD}&=C^{(0)}_{1,\DD\DD}+C^{(0)}_{2,\DD\DD}\,\bm\Sigma_1\!\cdot\!\bm\Sigma_2\\
&\qquad+C^{(0)}_{3,\DD\DD}\,\Sigma_1^{ij}\Sigma_2^{ij}+C^{(0)}_{4,\DD\DD}\,\Sigma_1^{ijk}\Sigma_2^{ijk},
\end{aligned}
\label{eq:ctLO}
\end{equation}
where $1$ and $2$ label the two baryons of the channel. Reference~\cite{sk2020} does not display the isospin factors of the transition terms, which at fixed total isospin are numbers that can be absorbed into the couplings. We keep them explicit because the large-$\Nc$ analysis of Sec.~\ref{sec:largeN} refers to the operators. The $\NN\to\ND$ term does not contribute in any S-wave, since the $\NN$ S-waves have $(S,I)=(0,1)$ or $(1,0)$ whereas $\ND$ has $S,I\in\{1,2\}$~\cite{sk2020}. The remaining terms carry $2+2+2+2+4=12$ couplings.

Equation~\eqref{eq:ctLO} contains no isospin dependence in the diagonal $\ND$ and $\DD$ terms. For $\NN$ this is no restriction, because an isospin-dependent operator is Fierz equivalent to the two displayed. For $\ND$ and $\DD$ it is a restriction. The two Pauli-allowed isospin values that accompany a given $S$ in $\DD$ scattering ($I=1,3$ for even $S$ and $I=0,2$ for odd $S$) and the two isospin values of $\ND$ are separated only by an isospin-dependent operator. The minimal completion of Eq.~\eqref{eq:ctLO} in S-waves adds
\begin{equation}
\begin{aligned}
V^{(0)}_{\mathrm{ct},\ND\ND}&\ni\big[\widetilde C^{(0)}_{1,\ND\ND}+\widetilde C^{(0)}_{2,\ND\ND}\,\bm\sigma_1\!\cdot\!\bm\Sigma_2\big]\bm\tau_1\!\cdot\!\bm\Theta_2,\\
V^{(0)}_{\mathrm{ct},\DD\DD}&\ni\Big[\sum_{n=1}^4\widetilde C^{(0)}_{n,\DD\DD}\,O_n\Big]\bm\Theta_1\!\cdot\!\bm\Theta_2,
\end{aligned}
\label{eq:ctLOiso}
\end{equation}
where $O_n$ are the four spin operators of the $\DD$ line of Eq.~\eqref{eq:ctLO}. This adds six couplings, for a total of 18, which is the number of independent LO S-wave couplings of the $N$--$\Delta$ system (Sec.~\ref{sec:sectors}) and agrees with the count of Ref.~\cite{rss2025}. The couplings $\widetilde C^{(0)}_n$ enter every $\ND$ and $\DD$ sector, including $(0,1)$ and $(1,0)$, but only in the combinations $C^{(0)}_n+\langle\bm\Theta_1\!\cdot\!\bm\Theta_2\rangle_I\widetilde C^{(0)}_n$ of Tab.~\ref{tab:dict}. Their role is to make the sectors with $I\ge2$ independent as follows, with the twelve couplings of Eq.~\eqref{eq:ctLO} alone, six of the seven sector couplings of these five sectors would coincide with those of the $I\le1$ sectors with the same $S$. The twelve couplings of Eq.~\eqref{eq:ctLO} determine the eleven sector couplings of the five sectors with $I\le1$, and their twelfth combination, built from $C^{(0)}_{2,\ND\DD}$ and $C^{(0)}_{3,\ND\DD}$, enters only the $I=2$ sector $(1,2)$. Nucleon--nucleon scattering at LO involves only the six sector couplings of $(0,1)$ and $(1,0)$, which Eq.~\eqref{eq:ctLO} parametrizes completely, and this is the sense in which the basis of Ref.~\cite{sk2020} is complete for $\NN$ scattering.

At next-to-leading order (NLO) the contact potentials are quadratic in the momenta. With $\bm q=\bm p'-\bm p$ and $\bm k=\tfrac12(\bm p'+\bm p)$ for the incoming and outgoing center-of-mass momenta $\bm p$ and $\bm p'$, the nucleonic part reads~\cite{sk2020}
\begin{equation}
\begin{aligned}
V^{(2)}_{\mathrm{ct},\NN\NN}&=C^{(2)}_1q^2+C^{(2)}_2k^2+\big(C^{(2)}_3q^2+C^{(2)}_4k^2\big)\,\bm\sigma_1\!\cdot\!\bm\sigma_2\\
&\quad+C^{(2)}_5\,i(\bm\sigma_1+\bm\sigma_2)\!\cdot\!(\bm q\times\bm k)\\
&\quad+C^{(2)}_6\,\bm\sigma_1\!\cdot\!\bm q\,\bm\sigma_2\!\cdot\!\bm q+C^{(2)}_7\,\bm\sigma_1\!\cdot\!\bm k\,\bm\sigma_2\!\cdot\!\bm k,
\end{aligned}
\label{eq:ctNLO}
\end{equation}
and the terms with external $\Delta$'s have the same structures built from $\bm S^\dagger$, $\bm\Sigma$ and the rank-two tensors, namely $q^2$ and $k^2$ multiplying each momentum-independent structure of Eq.~\eqref{eq:ctLO}, the spin--orbit terms $i(\bm\sigma_1\pm\bm\Sigma_2)\!\cdot\!(\bm q\times\bm k)$ and $i\bm S_1^\dagger\!\cdot\!(\bm q\times\bm k)$, the tensor terms $\bm X_1\!\cdot\!\bm q\,\bm Y_2\!\cdot\!\bm q$ and $\bm X_1\!\cdot\!\bm k\,\bm Y_2\!\cdot\!\bm k$, and the rank-two contractions $q^iq^kX_1^{ij}Y_2^{kj}$ and $k^ik^kX_1^{ij}Y_2^{kj}$. Their explicit form is Eq.~(C.1) of Ref.~\cite{sk2020}, with $5$ ($\NN\ND$), $8$ ($\ND\ND$), $8$ ($\NN\DD$), $9$ ($\ND\DD$) and $15$ ($\DD\DD$) couplings, 45 in all, before the isospin completion analogous to Eq.~\eqref{eq:ctLOiso}. In this work the NLO contact terms enter in two places only. First, the regulator-induced effective range of the LO cutoff theory, Eq.~\eqref{eq:CLambda} below, is of the size of a $C^{(2)}$ term, which is the counting statement that fixes the order of the analysis. Second, the momentum-dependent parts of vector-meson exchange belong to the $C^{(2)}$ terms and not to the $C^{(0)}$ terms, which is used in Sec.~\ref{sec:saturation}. No NLO coupling is fitted or flowed here.

\subsection{Sector couplings}\label{sec:sectorcouplings}

A momentum-independent contact interaction connects S-waves only and conserves the total spin $S$ and isospin $I$. The channels $\NN$, $\ND$ and $\DD$ with the same $(S,I)$ and $L=0$ form what we call an $(S,I)$ sector, and in each sector the contact potential is a real symmetric matrix of numbers $C^{(S,I)}_{\alpha\beta}$ with $\alpha,\beta\in\{\NN,\ND,\DD\}$, which we call the sector couplings. They are the quantities that flow. Tab.~\ref{tab:dict} gives them in terms of the couplings of Eqs.~\eqref{eq:ctLO} and \eqref{eq:ctLOiso}, and Tab.~\ref{tab:factors} lists the spin and isospin factors, which are the matrix elements of the operators in the coupled two-baryon states. We have computed these factors with the matrices of App.~\ref{app:conv}, and the diagonal elements and the $\ND\to\DD$ elements agree with the isospin factors of the one-pion-exchange potentials of Ref.~\cite{sk2020}. For $\NN\to\DD$ we use the operator eigenvalues as defined here. The corresponding entries of Ref.~\cite{sk2020} are one half of these values, which we attribute to a different normalization of the transition potential between identical-particle states and which does not affect the analysis, since only products of the transition coupling with its own spin--isospin factor enter.

The transition couplings are matrix elements between normalized channel states. For $\NN\to\DD$ both coupled states are antisymmetric as they stand and the matrix element is the product-state element of Tab.~\ref{tab:factors}. For $\ND\to\DD$ the single $\ND$ channel used here is the combination of the $\ND$ and $\Delta N$ channels of Ref.~\cite{sk2020} with the exchange symmetry of the $\DD$ state, and its coupling to $\DD$ is $\sqrt2$ times the product-state element, as derived in App.~\ref{app:conv}. This factor is included in Tab.~\ref{tab:dict}. The potentials of Ref.~\cite{sk2020} are defined with the opposite overall sign to ours, as we infer from their $K$-matrix relation, in which a positive potential gives a positive phase shift, and from the sign of their one-pion-exchange potential (App.~\ref{app:conv}). With identical operators, states and integration measure, the dictionary is
\begin{equation}
C_X^{\text{here}}=-\,C_X^{\text{Ref.~\cite{sk2020}}}
\label{eq:signdict}
\end{equation}
for every coupling $X$.

\begin{table}[t]
\caption{Sector couplings $C^{(S,I)}_{\alpha\beta}$ in terms of the contact couplings of Eqs.~\eqref{eq:ctLO} and \eqref{eq:ctLOiso}. $\langle O\rangle_S$ and $\langle O\rangle_I$ are the matrix elements of Tab.~\ref{tab:factors} for the $S$ and $I$ of the sector. The sign of the transition couplings depends on the phase convention of the $\Delta$ states, fixed here by the matrices of App.~\ref{app:conv} and Condon--Shortley coupling.}
\label{tab:dict}
\begin{ruledtabular}\footnotesize\setlength{\tabcolsep}{3pt}
\begin{tabular}{ll}
Sector coupling & Expression\\
\hline
$C^{(S,I)}_{\NN\NN}$ & $C_S+\langle\bm\sigma_1\!\cdot\!\bm\sigma_2\rangle_S\,C_T$\\[3pt]
$C^{(S,I)}_{\ND\ND}$ & $C^{(0)}_{1}+\langle\bm\sigma_1\!\cdot\!\bm\Sigma_2\rangle_S\,C^{(0)}_{2}$\\
 & $\ +\langle\bm\tau_1\!\cdot\!\bm\Theta_2\rangle_I\big[\widetilde C^{(0)}_1+\langle\bm\sigma_1\!\cdot\!\bm\Sigma_2\rangle_S \widetilde C^{(0)}_2\big]$\\[3pt]
$C^{(S,I)}_{\DD\DD}$ & $\sum_{n=1}^4\big[C^{(0)}_n+\langle\bm\Theta_1\!\cdot\!\bm\Theta_2\rangle_I \widetilde C^{(0)}_n\big]\,\langle O_n\rangle_S$\\
 & $O_n=\bone,\ \bm\Sigma_1\!\cdot\!\bm\Sigma_2,\ \Sigma_1^{ij}\Sigma_2^{ij},\ \Sigma_1^{ijk}\Sigma_2^{ijk}$\\[3pt]
$C^{(S,I)}_{\NN\DD}$ & $\langle\bm T_1^\dagger\!\cdot\!\bm T_2^\dagger\rangle_I\big[\langle\bm S_1^\dagger\!\cdot\!\bm S_2^\dagger\rangle_S C^{(0)}_{2}+\langle S_1^{ij\dagger}S_2^{ij\dagger}\rangle_S C^{(0)}_{3}\big]$\\[3pt]
$C^{(S,I)}_{\ND\DD}$ & $\sqrt2\,\langle\bm T_1^\dagger\!\cdot\!\bm\Theta_2\rangle_I\big[\langle\bm S_1^\dagger\!\cdot\!\bm\Sigma_2\rangle_S C^{(0)}_{2}+\langle S_1^{ij\dagger}\Sigma_2^{ij}\rangle_S C^{(0)}_{3}\big]$\\
\end{tabular}
\end{ruledtabular}
\end{table}

\begin{table}[t]
\caption{Matrix elements of the spin operators of Eq.~\eqref{eq:ctLO} in the coupled two-baryon spin states of total spin $S$, and of the isospin operators (in parentheses) in the states of total isospin $I$, which take the same values with $S\to I$. Transition elements are between the states of equal $S$ (or $I$) in the initial and final channel.}
\label{tab:factors}
\begin{ruledtabular}\footnotesize\setlength{\tabcolsep}{3pt}
\begin{tabular}{llcccc}
Operator & Channel & $0$ & $1$ & $2$ & $3$\\
\hline
$\bm\sigma_1\!\cdot\!\bm\sigma_2$ ($\bm\tau_1\!\cdot\!\bm\tau_2$) & $\NN$ & $-3$ & $1$ & -- & --\\
$\bm\sigma_1\!\cdot\!\bm\Sigma_2$ ($\bm\tau_1\!\cdot\!\bm\Theta_2$) & $\ND$ & -- & $-5$ & $3$ & --\\
$\bm\Sigma_1\!\cdot\!\bm\Sigma_2$ ($\bm\Theta_1\!\cdot\!\bm\Theta_2$) & $\DD$ & $-15$ & $-11$ & $-3$ & $9$\\
$\Sigma_1^{ij}\Sigma_2^{ij}$ & $\DD$ & $\tfrac{15}{2}$ & $\tfrac32$ & $-\tfrac92$ & $\tfrac32$\\
$\Sigma_1^{ijk}\Sigma_2^{ijk}$ & $\DD$ & $-\tfrac{350}{3}$ & $70$ & $-\tfrac{70}{3}$ & $\tfrac{10}{3}$\\
$\bm S_1^\dagger\!\cdot\!\bm S_2^\dagger$ ($\bm T_1^\dagger\!\cdot\!\bm T_2^\dagger$) & $\NN\to\DD$ & $-\sqrt2$ & $-\tfrac{\sqrt{10}}{3}$ & -- & --\\
$S_1^{ij\dagger}S_2^{ij\dagger}$ & $\NN\to\DD$ & $\tfrac{5\sqrt2}{3}$ & $-\tfrac{\sqrt{10}}{3}$ & -- & --\\
$\bm S_1^\dagger\!\cdot\!\bm\Sigma_2$ ($\bm T_1^\dagger\!\cdot\!\bm\Theta_2$) & $\ND\to\DD$ & -- & $-2\sqrt{\tfrac53}$ & $-2\sqrt3$ & --\\
$S_1^{ij\dagger}\Sigma_2^{ij}$ & $\ND\to\DD$ & -- & $-\sqrt{10}$ & $\sqrt2$ & --\\
\end{tabular}
\end{ruledtabular}
\end{table}

\subsection{The S-wave sectors}\label{sec:sectors}

For $\NN$ and $\DD$ the Pauli principle with $L=0$ requires $S+I$ odd, whereas $\ND$ has $S,I\in\{1,2\}$ without constraint. Enumerating $(S,I)$ gives the ten sectors of Tab.~\ref{tab:blocks}, and the number of independent sector couplings is $\sum_bn_b(n_b+1)/2=18$. Two features of Tab.~\ref{tab:blocks} organize the rest of this work. First, the $\ND$ channel never couples to $\NN$ in an S-wave. It appears alone in $(1,1)$ and $(2,2)$ and together with $\DD$ in $(1,2)$ and $(2,1)$. Its first coupling to $\NN$ is the two-derivative transition $\NN\,\pw{1}{D}{2}\leftrightarrow\ND\,\pw{5}{S}{2}$ in $I=1$, so that the $J^P=2^+$, $I=1$ system is the first genuine three-channel problem, which we treat in App.~\ref{app:wilson}. Second, below the $\ND$ threshold only the two $\NN$ sectors contain an open channel. The six remaining sectors are closed, and their ten couplings are boundary data in the pionless regime (Sec.~\ref{sec:frozen}).

We state explicitly how the three types of $\ND$ transitions are treated. The $\NN\to\ND$ contact term of Eq.~\eqref{eq:ctLO} vanishes in S-waves by quantum numbers alone, so that at LO the absence of $\NN$--$\ND$ coupling is exact. The first $\NN\leftrightarrow\ND$ couplings are the NLO terms of order $q^2$, namely the transition $\NN\,\pw{1}{D}{2}\leftrightarrow\ND\,\pw{5}{S}{2}$ and the $P$-wave transitions $\NN\,\pw{3}{P}{J}\leftrightarrow\ND\,\pw{3,5}{P}{J}$, all in $I=1$. This is why the $\Delta$ contact terms reach every $\NN$ partial wave with $\ell\le3$ except $\pw{3}{D}{2}$, which has $I=0$~\cite{sk2020}, and it is also the limitation of the present LO analysis for $\NN$ scattering: the $\ND$ channel first enters the $\NN$ observables at NLO in the $P$ and $D$ waves, which we do not flow. The $\ND\to\ND$ and $\ND\to\DD$ couplings, by contrast, are fully part of the analysis. They occupy the four sectors $(1,1)$, $(2,2)$, $(1,2)$ and $(2,1)$ with eight of the 18 couplings, and they enter the large-$\Nc$ relations and their flow in Secs.~\ref{sec:largeN} and \ref{sec:twocut}. Below $E=\dM$ every channel of these sectors is closed, so that their couplings are boundary data in the pionless regime (Sec.~\ref{sec:frozen}), and the detailed flow analysis of Sec.~\ref{sec:decoupling} concerns the two sectors with an open channel. The regime in which the $\ND$ sectors cease to be frozen data, $\Lambda\sim\kND$ and $E\to\dM$, is the $\ND$ threshold region seen in the $\NN$ $\pw{1}{D}{2}$ wave at $T_{\rm lab}\simeq630$~MeV, outside pionless EFT and in the domain of $\Delta$-full chiral calculations with cutoffs of $450$--$700$~MeV~\cite{sk2020}.

\begin{table}[t]
\caption{The S-wave $(S,I)$ sectors of the $N$--$\Delta$ two-baryon system at LO, the channels they contain, the number $n_C$ of independent couplings and their status for energies below the $\ND$ threshold. The last column marks the sectors whose couplings are determined by the twelve couplings of Eq.~\eqref{eq:ctLO} alone ($C$) and those that become independent of the $I\le1$ sectors only through the isospin completion of Eq.~\eqref{eq:ctLOiso} ($\widetilde C$).}
\label{tab:blocks}
\begin{ruledtabular}\small
\begin{tabular}{llccc}
$(S,I)$ & Channels & $n_C$ & $E<\dM$ & \\
\hline
$(0,1)$ & $\NN\,\pw{1}{S}{0}$, $\DD\,\pw{1}{S}{0}$ & 3 & $\NN$ open & $C$\\
$(1,0)$ & $\NN\,\pw{3}{S}{1}$, $\DD\,\pw{3}{S}{1}$ & 3 & $\NN$ open & $C$\\
$(1,1)$ & $\ND\,\pw{3}{S}{1}$ & 1 & closed & $C$\\
$(2,1)$ & $\ND\,\pw{5}{S}{2}$, $\DD\,\pw{5}{S}{2}$ & 3 & closed & $C$\\
$(3,0)$ & $\DD\,\pw{7}{S}{3}$ & 1 & closed & $C$\\
$(1,2)$ & $\ND\,\pw{3}{S}{1}$, $\DD\,\pw{3}{S}{1}$ & 3 & closed & $C$, $\widetilde C$\\
$(2,2)$ & $\ND\,\pw{5}{S}{2}$ & 1 & closed & $\widetilde C$\\
$(0,3)$ & $\DD\,\pw{1}{S}{0}$ & 1 & closed & $\widetilde C$\\
$(2,3)$ & $\DD\,\pw{5}{S}{2}$ & 1 & closed & $\widetilde C$\\
$(3,2)$ & $\DD\,\pw{7}{S}{3}$ & 1 & closed & $\widetilde C$\\
\hline
 & total & 18 & & \\
\end{tabular}
\end{ruledtabular}
\end{table}

%% ================================================================
\section{Cutoff flow of the coupled channels and decoupling}\label{sec:decoupling}
%% ================================================================

\subsection{Single channel}\label{sec:single}

We first recall the LO cutoff theory for one open channel with a momentum- and energy-independent coupling $C(\Lambda)$, which fixes the normalization used throughout. The loop integral of Eq.~\eqref{eq:LS} is, for $0<k<\Lambda$ (App.~\ref{app:int}),
\begin{equation}
B(E;\Lambda)=-\frac{\mu}{\pi^2}\Big[\Lambda-\frac{k}{2}\ln\frac{\Lambda+k}{\Lambda-k}\Big]-\frac{i\mu k}{2\pi},\qquad k=\sqrt{2\mu E}.
\label{eq:Iopen}
\end{equation}
With $\mathcal A^{-1}=B-1/C$ and $\mathcal A^{-1}=(\mu/2\pi)(k\cot\delta-ik)$ the imaginary parts match, which is two-body unitarity, and the real parts give
\begin{equation}
k\cot\delta=-\frac{2\Lambda}{\pi}+\frac{k}{\pi}\ln\frac{\Lambda+k}{\Lambda-k}-\frac{2\pi}{\mu\,C(\Lambda)},
\label{eq:kcot}
\end{equation}
where $\frac{k}{\pi}\ln\frac{\Lambda+k}{\Lambda-k}=\frac{2k^2}{\pi\Lambda}+\frac{2k^4}{3\pi\Lambda^3}+\mathcal O(k^6/\Lambda^5)$. Matching to the effective-range expansion at threshold, one obtains
\begin{equation}
C(\Lambda)=\frac{2\pi}{\mu}\,\frac{1}{1/a-2\Lambda/\pi},\quad
r_\Lambda=\frac{4}{\pi\Lambda},\quad
v_{2,\Lambda}=\frac{2}{3\pi\Lambda^3}.
\label{eq:CLambda}
\end{equation}
With $\hat C\equiv\mu\Lambda C/\pi^2$ the flow is $\Lambda\,\dd\hat C/\dd\Lambda=\hat C+\hat C^2$, with the trivial fixed point $\hat C=0$ and the unitary fixed point $\hat C=-1$, that is $C=-\pi^2/(\mu\Lambda)$. It has the form of the PDS flow of Ref.~\cite{rss2023} under the identification $\mu_{\rm PDS}=2\Lambda/\pi$, which we use whenever a comparison with that scheme is made. The LO cutoff theory has a nonzero effective range, $r_\Lambda=2.51$, $1.67$ and $0.84$~fm at $\Lambda=100$, $150$ and $300$~MeV, against the physical values $2.75$~fm ($\pw{1}{S}{0}$) and $1.75$~fm ($\pw{3}{S}{1}$). It is a regulator artifact of the size of the NLO operator. For this reason a LO cutoff theory reproduces the physical effective ranges for $\Lambda\simeq90$--$145$~MeV, and the residual cutoff dependence of LO observables scales as $1/\Lambda$~\cite{gh2020}. In PDS the LO effective range vanishes, so that the two schemes agree on $1/a$ and differ on $r$ by $r_\Lambda$ at this order. For $\Lambda\to\infty$ at fixed $a$, $C\to-\pi^2/(\mu\Lambda)\to0^-$ and $r_\Lambda\to0^+$, consistent with the Wigner bound for a zero-range interaction~\cite{pbc1998,bcp1998}. A numerical solution of Eq.~\eqref{eq:LS} with Eq.~\eqref{eq:CLambda} on a momentum mesh reproduces $1/a$ to $10^{-4}$ and $r_\Lambda$ to $0.3\%$.

\subsection{The closed-channel loop}\label{sec:bubble}

We now turn to the two sectors $(0,1)$ and $(1,0)$ in which the $\NN$ channel (channel 1) couples to $\DD$ (channel 2). Below the $\DD$ threshold the loop integral of channel 2 is, from Eq.~\eqref{eq:LS} with a sharp cutoff (App.~\ref{app:int}),
\begin{equation}
B_2(E;\Lambda)=-\frac{\mu_2}{\pi^2}\Big[\Lambda-\kappa\arctan\frac{\Lambda}{\kappa}\Big],\quad
\kappa=\sqrt{2\mu_2(2\dM-E)},
\label{eq:Iclosed}
\end{equation}
with $\kappa(0)=\kDD$. It is the analytic continuation $k\to i\kappa$ of Eq.~\eqref{eq:Iopen}. Its two limits are
\begin{equation}
\begin{aligned}
B_2&\simeq-\frac{\mu_2}{\pi^2}\,\frac{\Lambda^3}{3\kappa^2}\Big(1-\frac{3\Lambda^2}{5\kappa^2}\Big)\qquad(\Lambda\ll\kappa),\\
B_2&\simeq-\frac{\mu_2}{\pi^2}\Lambda+\frac{\mu_2\kappa}{2\pi}-\frac{\mu_2\kappa^2}{\pi^2\Lambda}\qquad(\Lambda\gg\kappa).
\end{aligned}
\label{eq:Ilimits}
\end{equation}
The large-$\Lambda$ form reproduces the PDS expression $(\mu_2/2\pi)(\kappa-\mu_{\rm PDS})$ with $\mu_{\rm PDS}=2\Lambda/\pi$, which is mass independent and keeps the finite piece $\mu_2\kappa/2\pi$ of the closed channel. The small-$\Lambda$ form is the decoupling: the closed channel contributes a term of order $\Lambda^3/\kappa^2$ instead of the linear divergence, which relative to the open loop $-\mu_1\Lambda/\pi^2$ is a correction of relative order $\Lambda^2/\kappa^2$. The crossover between the two regimes is governed by the function $R(x)=1-\arctan(x)/x$ with $x=\Lambda/\kDD$, which multiplies $-\mu_2\Lambda/\pi^2$ and interpolates between $x^2/3$ and $1-\pi/(2x)$. Its values are $R=0.0046$, $0.039$, $0.13$ and $0.19$ at $\Lambda=100$, $300$, $600$ and $770$~MeV, and the ratio of the closed to the open loop, $(\mu_2/\mu_1)R$, is $0.006$, $0.051$, $0.17$ and $0.25$ at the same cutoffs. Below $300$~MeV the $\DD$ loop is at the $5\%$ level of a linear divergence that is itself absorbed into the $\NN$ coupling, whereas at $770$~MeV it is one quarter of the open loop, so that at the vector-meson scale the $\DD$ channel is not decoupled.

\subsection{Exact flow and the decoupled regime}\label{sec:exactflow}

For a symmetric $2\times2$ matrix $C(\Lambda)$ of momentum-independent couplings, the requirement that the full matrix of on-shell amplitudes at $E=0$ be independent of $\Lambda$, $\partial_\Lambda[C^{-1}-B(0;\Lambda)]=0$, gives the exact LO flow
\begin{equation}
\Lambda\frac{\dd C}{\dd\Lambda}=C\,\mathcal D\,C,\qquad
\mathcal D=\mathrm{diag}\Big(\frac{\mu_1\Lambda}{\pi^2},\ \frac{\mu_2}{\pi^2}\frac{\Lambda^3}{\Lambda^2+\kDD^2}\Big),
\label{eq:flowC}
\end{equation}
whose solution is
\begin{equation}
C^{-1}(\Lambda)=K+B(0;\Lambda)
\label{eq:exactsol}
\end{equation}
for a constant symmetric matrix $K$, the negative inverse of the threshold amplitude matrix, $K=-\mathcal A^{-1}(E=0)$, which we call the threshold matrix. This renormalization condition includes the closed-channel elements $K_{12}$ and $K_{22}$, which no $\NN$ datum constrains, and it is in this sense a choice: it defines the two-channel contact theory whose couplings are followed. The $\NN$ scattering length fixes one combination of its elements, $K_{11}-K_{12}^2/K_{22}=\mu_1/(2\pi a)$. Equation~\eqref{eq:exactsol} holds for any value of $\Lambda/\kDD$ and is the basis of everything that follows. In the decoupled regime $\Lambda\ll\kDD$ the flow of Eq.~\eqref{eq:flowC} reads, with $x_1\equiv\mu_1\Lambda/\pi^2$,
\begin{equation}
\begin{aligned}
\Lambda\,\frac{\dd C_{11}}{\dd\Lambda}&=x_1C_{11}^2\left[1+\mathcal O\left(\frac{\Lambda^2}{\kDD^2}\right)\right],\\
\Lambda\,\frac{\dd C_{12}}{\dd\Lambda}&=x_1C_{11}C_{12}\left[1+\mathcal O\left(\frac{\Lambda^2}{\kDD^2}\right)\right],\\
\Lambda\,\frac{\dd C_{22}}{\dd\Lambda}&=x_1C_{12}^2\left[1+\mathcal O\left(\frac{\Lambda^2}{\kDD^2}\right)\right],
\end{aligned}
\label{eq:fdec}
\end{equation}
which is solved by $C_{11}(\Lambda)$ of Eq.~\eqref{eq:CLambda} together with
\begin{equation}
\eta\equiv\frac{C_{12}}{C_{11}}=\text{const},\qquad
\bar C_{22}\equiv C_{22}-\frac{C_{12}^2}{C_{11}}=\text{const},
\label{eq:inv}
\end{equation}
up to corrections of relative order $\Lambda^2/\kDD^2$. The combination $\bar C_{22}$ is the Schur complement of the $\NN$ entry in the coupling matrix, that is, the effective $\DD$ coupling obtained when the $\NN$ channel is eliminated at the level of the potential, $C_{22}-C_{12}C_{11}^{-1}C_{12}$, and its inverse is the $\DD$ element of $C^{-1}$. Equivalently, $(C^{-1})_{12}=-\eta/\bar C_{22}$ and $(C^{-1})_{22}=1/\bar C_{22}$ are, in this regime, the $\Lambda$-independent elements $K_{12}$ and $K_{22}$ of the threshold matrix, whereas $(C^{-1})_{11}=1/C_{11}+\eta^2/\bar C_{22}$ runs through $C_{11}$ only. The exact statement is Eq.~\eqref{eq:exactsol}: the data that the $\Delta$ channels transmit from one cutoff to another are $K_{12}$ and $K_{22}$, and the transition ratio and the Schur complement depend on $\Lambda$ only through the closed loop,
\begin{equation}
\eta(\Lambda)=-\frac{K_{12}}{K_{22}+B_2(0;\Lambda)},\qquad
\bar C_{22}(\Lambda)=\frac{1}{K_{22}+B_2(0;\Lambda)}.
\label{eq:etaL}
\end{equation}
Their infrared limits
\begin{equation}
\eta_K\equiv-\frac{K_{12}}{K_{22}},\qquad \bar C_K\equiv\frac{1}{K_{22}}
\label{eq:etaK}
\end{equation}
are exact RG invariants, and we call them the invariants of the $\Delta$ channels. For $\Lambda\ll\kDD$ the running quantities $\eta(\Lambda)$ and $\bar C_{22}(\Lambda)$ approach them with corrections of relative order $\Lambda^2/\kDD^2$, and the ratio is
\begin{equation}
\frac{\eta_K}{\eta(\Lambda)}=\frac{\bar C_K}{\bar C_{22}(\Lambda)}=\frac{1}{1-B_2(0;\Lambda)\,\bar C_{22}(\Lambda)} .
\label{eq:etaratio}
\end{equation}
At $\Lambda=770$~MeV, where the closed loop is one quarter of the open one, these corrections are of order one: for the boundary conditions of Sec.~\ref{sec:flow} the ratio of Eq.~\eqref{eq:etaratio} lies between $1.08$ and $1.71$, against $1.01$ to $1.05$ when the boundary condition is imposed at $200$~MeV (Tab.~\ref{tab:family}). We therefore quote $\eta_K$ and $\bar C_K$, equivalently $K_{12}$ and $K_{22}$, as the transmitted data and reserve $\eta(\Lambda)$ and $\bar C_{22}(\Lambda)$ for $\Lambda\ll\kDD$. The content of Eq.~\eqref{eq:inv} is that the closed loop is small there, so that the decoupling of the $\DD$ channel is automatic in the sense of Appelquist and Carazzone~\cite{ac1975}. The transition coupling itself is not frozen in that regime. It runs like $C_{11}$, because every $\NN\to\DD$ transition is dressed by $\NN$ rescattering, whereas the ratio $\eta$ and the Schur complement $\bar C_{22}$, which is the $\DD$ coupling with the $\NN$ channel integrated out, are.

\subsection{Matching correction}\label{sec:matchcorr}

\begin{figure*}[t]
\includegraphics[width=\textwidth]{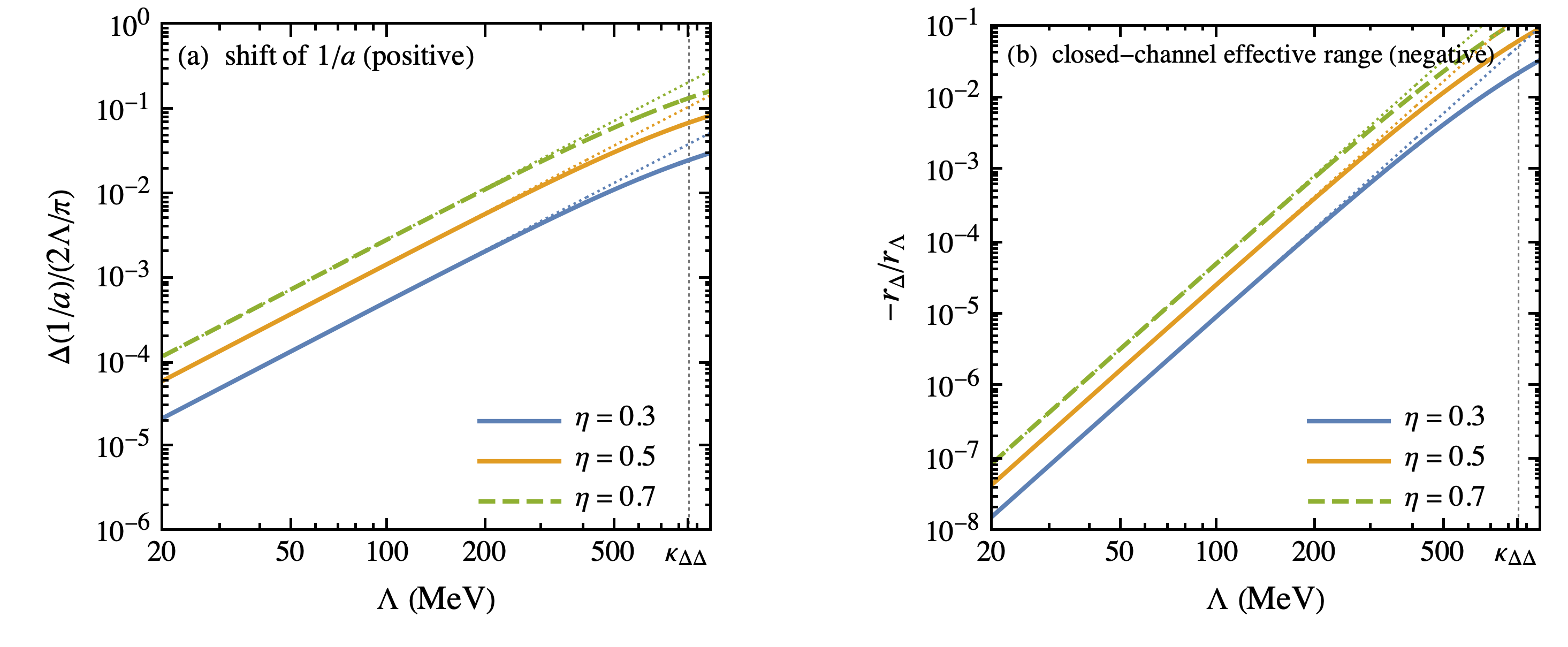}
\caption{Decoupling law. (a) Shift of $1/a$ between the two-channel theory and the nucleon-only theory with the same $\NN$ coupling, relative to $2\Lambda/\pi$, for the transition ratios $\eta=0.3$, $0.5$ and $0.7$. Solid and dashed lines use the exact closed-channel function $F(\Lambda)$ of Eq.~\eqref{eq:Iclosed}, dotted lines its leading form $\Lambda^3/(3\kDD^2)$. (b) The closed-channel effective range, with its sign reversed, relative to the cutoff-induced $r_\Lambda=4/(\pi\Lambda)$, with the same line styles. The vertical dotted line is $\Lambda=\kDD$.}
\label{fig:dec}
\end{figure*}

The decoupling can be stated as a matching between the two-channel theory and the nucleon-only theory with the same $\NN$ coupling. Eliminating channel 2 at the level of the potential, $C_{\rm eff}(E)=C_{11}+C_{12}^2B_2(E)/[1-C_{22}B_2(E)]$, and expanding to first order in the closed-channel loop, the shifts of the $\NN$ threshold parameters, defined as the two-channel values minus the nucleon-only values, are
\begin{equation}
\begin{aligned}
\Delta\Big(\frac1a\Big)&=+\frac{2\pi}{\mu_1}\,\eta^2\,\frac{\mu_2\Lambda^3}{3\pi^2\kDD^2}\Big[1+\mathcal O\Big(\frac{\Lambda^2}{\kDD^2}\Big)\Big],\\
r_\Delta&=-\frac{4}{3\pi}\,\frac{\mu_2^2}{\mu_1^2}\,\eta^2\,\frac{\Lambda^3}{\kDD^4}\Big[1+\mathcal O\Big(\frac{\Lambda^2}{\kDD^2}\Big)\Big],
\end{aligned}
\label{eq:shifts}
\end{equation}
and the exact forms follow from Eq.~\eqref{eq:exactsol}, $\Delta(1/a)=(2\pi/\mu_1)K_{12}^2[1/(K_{22}+B_2)-1/K_{22}]$. The shift of $1/a$ is positive, since the coupled closed channel adds attraction, and the induced effective range is negative. Tab.~\ref{tab:num2} compares the exact and the leading forms for a transition coupling of natural size, $|C_{12}|\sim2\pi/(\mu_2\Lambda)$, and Fig.~\ref{fig:dec} shows them as functions of $\Lambda$ for fixed $\eta$. The shift of $1/a$ is not observable, since $C_{11}$ is fitted to $a$, but it measures the size of the matching correction between the $\Delta$-full and the nucleon-only theory. Relative to the bare inverse coupling $|1/a_s-2\Lambda/\pi|$ it is $0.2\%$, $1.3\%$ and $4.3\%$ at $\Lambda=100$, $300$ and $600$~MeV. Relative to $|1/a_s|$ it is $1.4\%$ and $30\%$ at $100$ and $300$~MeV, the large number reflecting the fine tuning of the $\pw{1}{S}{0}$ channel rather than a large closed-channel effect. The $\DD$-induced effective range is more than two orders of magnitude below the regulator artifact $r_\Lambda$. The closed channel is thus invisible in the $\NN$ observables at this order, which is our interpretation of the numerical insensitivity found in Ref.~\cite{sk2020}. Their statement is numerical and model specific, whereas Eq.~\eqref{eq:shifts} is kinematic.

The attribution of the effective range to the closed channel is scheme dependent. In PDS the $\Delta$-induced effective range is $r_\Delta=-(\mu_2/\mu_1)\Kh_{12}^2/[\kDD(\Kh_{22}-\kDD)^2]$ in the notation of Ref.~\cite{rss2023} and is not suppressed by powers of $\Lambda/\kDD$, because PDS is mass independent and keeps the finite piece of Eq.~\eqref{eq:Ilimits}, as does the $\overline{\rm MS}$ treatment of the $\DD$ channel in Ref.~\cite{savage1997}. In the cutoff scheme with $\Lambda<\kDD$ essentially all of the effective range sits in the two-derivative $\NN$ coupling. There is no contradiction, since the physical effective range is one number and only its sum is observable. Only external information on $\eta_K$ and $\bar C_K$ turns the $\Delta$-induced effective range into a meaningful quantity, and then it has to be computed in one scheme consistently.

\begin{table}[t]
\caption{Effect of the closed $\DD$ channel on the $\NN$ $\pw{1}{S}{0}$ threshold parameters for couplings of natural size, $C_{12}=C_{22}=-2\pi/(\mu_2\Lambda)$, compared with the leading forms of Eq.~\eqref{eq:shifts}. The shifts are the two-channel values minus the nucleon-only values at the same $C_{11}$.}
\label{tab:num2}
\begin{ruledtabular}\footnotesize\setlength{\tabcolsep}{3pt}
\begin{tabular}{cccccc}
$\Lambda$ & $(\Lambda/\kDD)^3$ & $\Delta(1/a)$ & leading & $r_\Delta$ & leading\\
(MeV) & & (MeV) & (MeV) & (fm) & (fm)\\
\hline
100 & $1.6\times10^{-3}$ & $+0.115$ & $+0.116$ & $-0.00008$ & $-0.00008$\\
150 & $5.5\times10^{-3}$ & $+0.357$ & $+0.362$ & $-0.00025$ & $-0.00026$\\
300 & $4.4\times10^{-2}$ & $+2.52$ & $+2.67$ & $-0.0017$ & $-0.0019$\\
600 & $3.5\times10^{-1}$ & $+16.6$ & $+20.5$ & $-0.0098$ & $-0.0147$\\
\end{tabular}
\end{ruledtabular}
\end{table}

\subsection{Fixed points in the large-threshold regime}\label{sec:fixedpoints}

The flow can be placed in the fixed-point language of Refs.~\cite{bmr1999,lensky2011,cgvk2004}. Allowing the potential to depend on the energy and demanding RG invariance of the full off-shell amplitude, the flow takes the Wilsonian form of App.~\ref{app:wilson} in the rescaled variables $\Vh=(\Lambda/\pi^2)M^{1/2}VM^{1/2}$, $\kh_i=k_i/\Lambda$ and $\kaph=\kDD/\Lambda$, with $M=\mathrm{diag}(\mu_1,\mu_2)$. For one channel it is the equation of Ref.~\cite{bmr1999}, and the new feature is the second scaling variable $\kaph$, which grows toward the infrared. In the regime $\Lambda>\kDD$ the equation has three types of analytic fixed points: the trivial point, a rank-two point at which both channels sit at the unitary limit, and a continuous family of rank-one points
\begin{equation}
\Vh_*^{(1)}[n]=-\frac{n\,n^{\Tp}}{n_1^2\varphi(\kh_1^2)+n_2^2\varphi(\kh_2^2)},\qquad n=(\cos\theta,\sin\theta),
\label{eq:rank1}
\end{equation}
with $\varphi(x)=1-\sqrt x\,\mathrm{artanh}\sqrt x$, labeled by the eigenchannel angle $\theta$, as found by Lensky and Birse for a small threshold~\cite{lensky2011} and in the sharp-cutoff renormalization of Ref.~\cite{cgvk2004}. Since $\LH=770$~MeV lies below $\kDD=850$~MeV, this regime is not entered in the present analysis, and we record it only for two properties. Around the $\NN$-aligned rank-one point the transition coupling is exactly marginal, which is the fixed-point counterpart of the invariance of $\eta$ in Eq.~\eqref{eq:inv}, and the rank-one points form a continuous family in $\theta$ that the flow does not select: the eigenchannel angle is fixed by the boundary condition, not by the dynamics. In the physical regime $\Lambda<\kDD$ the $\NN$ channel flows to the single-channel fixed points of Ref.~\cite{bmr1999} with the $\DD$ channel as an irrelevant spectator, and the explicit dependence of the LO couplings on $\dM$ is a perturbation with the same scaling dimension as the effective range, that is, an NLO effect in the Wilsonian counting.

\subsection{Closed sectors and the complete LO structure}\label{sec:frozen}

For a sector in which every channel is closed below $E=\dM$, the flow of Eq.~\eqref{eq:flowC} generalizes to $\Lambda\,\dd C/\dd\Lambda=C\mathcal DC$ with $\mathcal D_{ii}=(\mu_i/\pi^2)\Lambda^3/(\Lambda^2+\kappa_i^2)$ for every channel $i$. For $\Lambda\ll\kappa_i$ the relative change of a coupling of natural size $|C|\sim2\pi/(\mu\Lambda)$ over a range $\Delta\Lambda\sim\Lambda$ is $(2/\pi)(\Lambda/\kappa_i)^2$, which is $2\%$ ($\DD$) to $5\%$ ($\ND$) at $\Lambda=150$~MeV and $1\%$ to $2\%$ at $100$~MeV. The ten couplings of the six closed sectors of Tab.~\ref{tab:blocks} are therefore frozen boundary data in the pionless regime, though not at the vector-meson scale, where $\LH/\kND=1.4$ and $\LH/\kDD=0.9$ and the same estimate gives changes of order one. They enter the $\NN$ observables only through terms suppressed by $\Lambda^2/\kappa^2$ and, for the sectors that reach $\NN$ at all, by additional powers $(p/\Lambda)^{2\ell}$ from the derivative transitions, so that they cannot be determined from $\NN$ data at any order in the pionless theory. Their sources are lattice QCD in the $\DD$ and $\ND$ channels and the large-$\Nc$ relations of the next section. The first sector in which $\ND$ reaches $\NN$, the $J^P=2^+$, $I=1$ system with $\NN\,\pw{1}{D}{2}$, $\ND\,\pw{5}{S}{2}$ and $\DD\,\pw{5}{S}{2}$, requires a two-derivative transition operator. We give its Wilsonian flow and fixed points in App.~\ref{app:wilson}, where we also show that for a transition coupling of natural size the $\ND$ channel shifts the $\pw{1}{D}{2}$ contact by $5\%$ at $\Lambda=300$~MeV, so that the $\ND$ channel is invisible in the $\NN$ D waves below that cutoff.

%% ================================================================
\section{Large-$\Nc$ relations and their renormalization-group flow}\label{sec:largeN}
%% ================================================================

\subsection{Hartree counting and time reversal}\label{sec:hartree}\label{sec:Ttheorem}

We now ask what the $1/\Nc$ expansion says about the 18 sector couplings. On the SU(4) multiplet of $N$ and $\Delta$, the spin, isospin and spin--flavor generators are $J^i$, $I^a$ and $G^{ia}$, normalized so that $G^{ia}=\sigma^i\tau^a/4$ on a single quark. Following Witten~\cite{witten1979} and Kaplan and Manohar~\cite{km1997}, the two-body Hamiltonian of two baryons in the multiplet is written in the Hartree form
\begin{equation}
H_2=\Nc\sum_{\mathcal O}c_{\mathcal O}\,\Nc^{-n(\mathcal O)}\,\mathcal O,
\label{eq:hartree}
\end{equation}
where $\mathcal O$ runs over the spin--isospin scalar products of $n(\mathcal O)$ one-body operators of baryons 1 and 2, symmetric under $1\leftrightarrow2$, and $c_{\mathcal O}=\mathcal O(1)$. The factor $\Nc^{1-n}$ is the Hartree estimate of an $n$-quark interaction. The two-body operators that appear are
\begin{equation}
\begin{aligned}
\Mop&\equiv4\,G_1\!\cdot\!G_2,\qquad \Jop\equiv J_1\!\cdot\!J_2,\qquad \Iop\equiv I_1\!\cdot\!I_2,\\
\Xop&\equiv J_1^iI_1^aG_2^{ia}+G_1^{ia}J_2^iI_2^a,\qquad \Sop\equiv J_1^2+J_2^2,
\end{aligned}
\label{eq:ops}
\end{equation}
and $\Bop\equiv\Jop+\Iop+\Mop$ is the exchange part of the two-baryon SU(4) Casimir, so that $c_1\bone+g\Bop$ is the general exact-SU(4) invariant. To evaluate the matrix elements for general $\Nc$, we construct the totally symmetric representation with $\Nc$ boxes as $\Nc$ bosons in the four quark states and compute every one- and two-body matrix element numerically for $\Nc=3$ to $13$ (App.~\ref{app:su4}). Three exact results follow. Within a multiplet of spin and isospin $j$, $G^{ia}=c_jJ^iI^a$ with $c_j=(\Nc+2)/[4j(j+1)]$, so that $c_N=(\Nc+2)/3$ and $c_\Delta=(\Nc+2)/15$, which at $\Nc=3$ are the quark-model values $\tfrac53$ and $\tfrac13$. The $N\to\Delta$ transition strength is $\sum|\langle\Delta|G^{ia}|N\rangle|^2=(\Nc-1)(\Nc+5)/8$. Hence the diagonal and the transition matrix elements of $G^{ia}$ all grow linearly with $\Nc$, whereas those of $J^i$ and $I^a$ do not. A term with $n_G$ factors of $G$ and $n_J+n_I$ factors of $J$ and $I$ therefore contributes
\begin{equation}
\Nc^{1-n}\Nc^{n_G}=\Nc^{1-(n_J+n_I)}
\label{eq:order}
\end{equation}
to a coupling, and the suppression relative to the LO is $\Nc^{-(n_J+n_I)}$. This is the rule of Ref.~\cite{km1997}, now shown to hold with the same powers for matrix elements between $\Delta$ states and for $N\leftrightarrow\Delta$ transitions. Tab.~\ref{tab:closed} gives the matrix elements of $\Mop$ in the ten sectors in closed form, together with those of $\Jop$ and $\Iop$, which are $\Nc$ independent. Two features are used repeatedly. $\Jop-\Iop=\tfrac12[S(S+1)-I(I+1)]$ is a multiple of the unit matrix in every sector, because $J_k^2=I_k^2$ on the multiplet, and it is the only operator in Eq.~\eqref{eq:ops} that is odd under the spin--isospin flip $(S,I)\to(I,S)$. The transition elements of $\Mop$ scale as $(\Nc-1)(\Nc+5)$ for $\NN\to\DD$ and as $(\Nc+2)\sqrt{(\Nc-1)(\Nc+5)}$ for $\ND\to\DD$, and the $\ND$ diagonal elements contain an exchange term of the former type.

\begin{table}[t]
\caption{Matrix elements of $\Mop=4G_1\!\cdot\!G_2$ in the S-wave sectors for general odd $\Nc$, with $p\equiv\Nc+2$ and $q\equiv(\Nc-1)(\Nc+5)$, and the $\Nc$-independent diagonal elements of $\Jop$ ($\Iop$ follows by the flip $S\leftrightarrow I$). Two values in a flip pair refer to its first and second sector. Transition signs follow the phase convention of App.~\ref{app:conv}.}
\label{tab:closed}
\begin{ruledtabular}\footnotesize\setlength{\tabcolsep}{3pt}
\begin{tabular}{llccc}
Sector & Element & $\Mop$ & $\Jop$ & $\Mop$ ($\Nc=3$)\\
\hline
$(0,1),(1,0)$ & $\NN$ & $-p^2/12$ & $-\tfrac34,\ \tfrac14$ & $-25/12$\\
 & $\DD$ & $11p^2/60$ & $-\tfrac{15}{4},\ -\tfrac{11}{4}$ & $55/12$\\
 & $\NN\to\DD$ & $\sqrt5\,q/12$ & $0$ & $4\sqrt5/3$\\
$(1,1)$ & $\ND$ & $5p^2/36-q/72$ & $-\tfrac54$ & $13/4$\\
$(2,2)$ & $\ND$ & $p^2/20-q/8$ & $\tfrac34$ & $-3/4$\\
$(1,2),(2,1)$ & $\ND$ & $-p^2/12+q/24$ & $-\tfrac54,\ \tfrac34$ & $-17/12$\\
 & $\DD$ & $11p^2/300$ & $-\tfrac{11}{4},\ -\tfrac34$ & $11/12$\\
 & $\ND\to\DD$ & $-\sqrt5\,p\sqrt q/15$ & $0$ & $-4\sqrt5/3$\\
$(0,3),(3,0)$ & $\DD$ & $-3p^2/20$ & $-\tfrac{15}{4},\ \tfrac94$ & $-15/4$\\
$(2,3),(3,2)$ & $\DD$ & $-3p^2/100$ & $-\tfrac34,\ \tfrac94$ & $-3/4$\\
\end{tabular}
\end{ruledtabular}
\end{table}

The counting of Eq.~\eqref{eq:order} is sharpened by a generalized time reversal. The physical time-reversal operator $T$ is antiunitary and reverses the spin, $TJ^iT^{-1}=-J^i$, but it leaves $I^3$ invariant, because the electric charge is even under time reversal, and it reverses only $I^2$ in the standard basis. The operator that reverses all three isospin generators is $\Tp\equiv T\,e^{i\pi I^2}$, the product of time reversal with an isospin rotation by $\pi$ about the 2-axis, in the same way as $T=e^{i\pi J^2}\mathcal K$ reverses the spin. It is antiunitary, it is a symmetry of any Hamiltonian that is invariant under time reversal and isospin, and it gives $\Tp J^i\Tp^{-1}=-J^i$, $\Tp I^a\Tp^{-1}=-I^a$ and $\Tp G^{ia}\Tp^{-1}=+G^{ia}$, because $G$ carries one spin and one isospin index. A spin--isospin scalar with $n_J$ spin and $n_I$ isospin factors and any number of $G$'s has $\Tp$ parity $(-1)^{n_J+n_I}$, and a static two-body potential has to be $\Tp$ even.\footnote{Only an antiunitary operator can reverse all three components of an angular-momentum algebra. If a unitary $U$ satisfied $UJ^iU^{-1}=-J^i$ for $i=1,2,3$, conjugating $[J^i,J^j]=i\epsilon^{ijk}J^k$ would give $[J^i,J^j]=-i\epsilon^{ijk}J^k$, a contradiction, whereas an antiunitary operator escapes it because it also maps $i\to-i$. Parity is unitary and leaves $J^i$, $I^a$ and $G^{ia}$ invariant, so it does not constrain the static spin--isospin structure, and charge conjugation maps the two-baryon states to two-antibaryon states and gives no constraint within the sector. The sign $(-1)^{n_J+n_I}$ can therefore come only from time reversal, and the isospin rotation in $\Tp$ is needed because the physical $T$ leaves $I^3$ invariant. The argument applies to the momentum-independent couplings only: with momenta, $\bm p$ is $\Tp$ odd as well, and operators with one $J$ or $I$ and one power of momentum, such as the spin--orbit term of Ref.~\cite{km1997}, are allowed at relative order $1/\Nc$.} Every operator of relative order $\Nc^{-(n_J+n_I)}$ with $n_J+n_I$ odd is therefore absent from the S-wave contact potential, and the corrections to the LO start at relative order $1/\Nc^2$ in all ten sectors, for the $\Delta$ channels as for $\NN$. For $\NN$ this is the known result of Ref.~\cite{km1997}, in which the central spin-dependent terms appear at relative order $1/\Nc^2$. The generalized time reversal identifies the reason and extends the statement to the $\Delta$ channels. We have verified it numerically with the only scalar that contains a single isospin factor, $\epsilon^{abc}\{I_1^a,G_1^{ib}\}G_2^{ic}+(1\leftrightarrow2)$, whose matrix elements in the real channel basis are purely imaginary at $\Nc=3$, $5$ and $7$, as those of a $\Tp$-odd operator have to be.

\subsection{LO relations, the two bases, and the first corrections}\label{sec:LOrel}\label{sec:bases}\label{sec:nnlo}

At LO only products of $G$'s appear. Kaplan and Manohar state that the most general LO central potential is a function of the single operator $G_1\!\cdot\!G_2$~\cite{km1997}, because $G^{ia}/\Nc$ becomes a commuting rotation matrix in the large-$\Nc$ limit and every scalar built from two such matrices is a function of their trace. We have verified this on the $N\oplus\Delta$ space: the only irreducible four-$G$ scalar that does not reduce by the Dashen--Jenkins--Manohar identities~\cite{djm1995} differs from a cubic polynomial in $\Mop$, taken as an operator on the full spin--flavor tower, by a remainder that falls as $\Nc^{-2}$, from $0.65$ at $\Nc=3$ to $0.034$ at $\Nc=13$ in relative norm, and that lies entirely in the space of operators with two $J$ or $I$ factors. The LO S-wave contact potential is thus
\begin{equation}
V_{\rm LO}=\Nc\,f(\Mop/\Nc^2)=\sum_\lambda v_\lambda\Pop_\lambda,
\label{eq:VLO}
\end{equation}
with $f$ an arbitrary function and $\Pop_\lambda$ the spectral projectors of $\Mop$ on the 18-dimensional coupling space. At $\Nc=3$ the matrix elements of Tab.~\ref{tab:closed} give, in the two-channel sectors,
\begin{equation}
\begin{aligned}
\Mop_{(0,1)}&=\Mop_{(1,0)}=\begin{pmatrix}-\tfrac{25}{12}&\tfrac{4\sqrt5}{3}\\[2pt]\tfrac{4\sqrt5}{3}&\tfrac{55}{12}\end{pmatrix},\\
\Mop_{(1,2)}&=\Mop_{(2,1)}=\begin{pmatrix}-\tfrac{17}{12}&-\tfrac{4\sqrt5}{3}\\[2pt]-\tfrac{4\sqrt5}{3}&\tfrac{11}{12}\end{pmatrix},
\end{aligned}
\label{eq:Mblocks}
\end{equation}
with eigenvalues $\tfrac54\pm2\sqrt5$ and $-\tfrac14\pm\tfrac{\sqrt{41}}2$, and $\Mop=\tfrac{13}4,-\tfrac34,-\tfrac{15}4,-\tfrac34$ in $(1,1)$, $(2,2)$, $(0,3)=(3,0)$ and $(2,3)=(3,2)$. The value $-\tfrac34$ is shared by $(2,2)$ and $(2,3)$, $(3,2)$. There are seven distinct eigenvalues, so that seven parameters determine the 18 couplings and the LO contains eleven relations, listed in Tab.~\ref{tab:relations}. Eight are flip relations $C^{(S,I)}_{\alpha\beta}=C^{(I,S)}_{\alpha\beta}$, because $\Mop$ is invariant under the automorphism $J\leftrightarrow I$, $G^{ia}\leftrightarrow G^{ai}$ that maps $(S,I)$ onto $(I,S)$. Two are eigenchannel relations, because the coupling matrix of a two-channel sector is a function of $\Mop$ and has its eigenvectors,
\begin{equation}
\begin{aligned}
\frac{C_{\DD}-C_{\NN}}{\Ctr}\bigg|_{(0,1),(1,0)}&=\sqrt5,\\
\frac{C_{\DD}-C_{\ND}}{\Ctr}\bigg|_{(1,2),(2,1)}&=-\frac{7}{4\sqrt5}.
\end{aligned}
\label{eq:eigrel}
\end{equation}
The last is accidental to $\Nc=3$, $C^{(2,2)}_{\ND}=C^{(2,3)}_{\DD}=C^{(3,2)}_{\DD}$, which is absent already at $\Nc=5$. In terms of the eigenchannel angle, the eigenvectors of $\Mop_{(0,1)}$ are $(\cos\theta,\sin\theta)$ with $\tan2\theta=2\Mop_{12}/(\Mop_{11}-\Mop_{22})=-2\sqrt5/5$, that is
\begin{equation}
\theta_{\mathrm A}=69.1^\circ\ (\DD\text{ probability }0.873),\quad \theta_{\mathrm A}-90^\circ=-20.9^\circ,
\label{eq:thetaA}
\end{equation}
and $\theta=34.3^\circ$, $-55.7^\circ$ in $(1,2)$. The first relation of Eq.~\eqref{eq:eigrel} is the statement $\theta=\theta_{\mathrm A}$, and it holds for any function $f$, not only for the linear truncation $V_{\mathcal A}=a_0\bone+a_G\Mop$ used in Refs.~\cite{rss2025,bvs2026}.

\begin{table}[t]
\caption{The eleven LO large-$\Nc$ relations among the 18 S-wave couplings at $\Nc=3$, their origin, and whether the LO cutoff flow transmits them unchanged from $\LH=770$~MeV to $\LL=140$~MeV (Secs.~\ref{sec:flipflow} and \ref{sec:degenerate}). The violations quoted for the last three rows are for boundary conditions of natural size with $|\bh|$, $|\bh'|$ between $0.05$ and $0.3$.}
\label{tab:relations}
\begin{ruledtabular}\scriptsize\setlength{\tabcolsep}{2pt}
\begin{tabular}{lll}
Relation & Origin & Transmitted (violation)\\
\hline
$C^{(0,1)}_{\alpha\beta}=C^{(1,0)}_{\alpha\beta}$ (3) & flip & exactly\\
$C^{(1,2)}_{\alpha\beta}=C^{(2,1)}_{\alpha\beta}$ (3) & flip & exactly\\
$C^{(0,3)}_{\DD}=C^{(3,0)}_{\DD}$ & flip & exactly\\
$C^{(2,3)}_{\DD}=C^{(3,2)}_{\DD}$ & flip & exactly\\
$(C_{\DD}-C_{\NN})/\Ctr=\sqrt5$, $(0,1)$ & eigenvectors & no ($\theta$: $3^\circ$--$45^\circ$)\\
$(C_{\DD}-C_{\ND})/\Ctr=-\frac{7}{4\sqrt5}$, $(1,2)$ & eigenvectors & no ($\theta$: $1^\circ$--$9^\circ$)\\
$C^{(2,2)}_{\ND}=C^{(2,3)}_{\DD}$ & accidental & no ($14\%$)\\
\end{tabular}
\end{ruledtabular}
\end{table}

Two two-parameter forms of the LO potential have been used in the literature. The Hartree basis $V_{\mathcal A}=a_0\bone+a_G\Mop$ is $f$ truncated to a linear function. The exact-SU(4) form $V_{\mathcal B}=C_{140}\Pop_{140}+C_{50}\Pop_{50}=c_1\bone+g\Bop$, which is the two-parameter reduction of Ref.~\cite{rss2025} and of our SU(3) analysis~\cite{bvs2026}, projects on the two SU(4) irreducible components $140\oplus50$ of the antisymmetric part of $20\otimes20$, on which $\Bop=\tfrac34$ and $-\tfrac{21}4$. Since $\Bop=\Mop+(\Jop+\Iop)$, the form $V_{\mathcal B}$ is $V_{\mathcal A}$ plus the operator $g(\Jop+\Iop)$, which is of relative order $1/\Nc^2$ by Eq.~\eqref{eq:order}, with its coefficient tied to that of $\Mop$ by exact SU(4) rather than left free. The two sets of projectors do not commute. In the $(0,1)$ sector,
\begin{equation}
\Bop_{(0,1)}=\begin{pmatrix}-\tfrac{31}{12}&\tfrac{4\sqrt5}{3}\\[2pt]\tfrac{4\sqrt5}{3}&-\tfrac{23}{12}\end{pmatrix},\qquad
\theta_{\mathrm B}=48.2^\circ,
\label{eq:Bblock}
\end{equation}
with $\DD$ probability $0.556$ against $0.873$ for $\theta_{\mathrm A}$. Both forms satisfy the eight flip relations, since $\Bop$ is flip invariant. $V_{\mathcal B}$ violates the eigenchannel relations, giving $(C_{\DD}-C_{\NN})/\Ctr=1/(2\sqrt5)$ in $(0,1)$ instead of $\sqrt5$, and it imposes relations that no function of $\Mop$ gives, such as $C^{(1,1)}_{\ND}=C^{(2,2)}_{\ND}$, because all channels in the 140 share one coupling. Neither form is the general LO with its seven parameters.

The size of the difference is not controlled by the counting. Tab.~\ref{tab:size} lists the diagonal matrix elements of $\Mop$ and of $\Jop+\Iop$ at $\Nc=3$. In the $\NN$ channels the ratio is $0.24$, close to the nominal $1/\Nc^2$ times a factor of two. In every channel that contains a $\Delta$ it lies between $0.40$ and $3.8$. The reason is that $\langle\Jop\rangle$ and $\langle\Iop\rangle$ for the $\Delta$ involve $j(j+1)=\tfrac{15}4$, so that $\langle\Jop+\Iop\rangle_{\DD}=-\tfrac{13}2$ in $(0,1)$ against $-\tfrac12$ for $\NN$, whereas $\langle\Mop\rangle$ grows only from $\tfrac{25}{12}$ to $\tfrac{55}{12}$. The expansion parameter for the $\Delta$ channels is $j(j+1)/\Nc^2$ with $j=\Nc/2$ at $\Nc=3$, which is of order one. The $(0,1)$ ratio $(C_{\DD}-C_{\NN})/\Ctr$ is $\tfrac{16}{5\sqrt5}(\Nc+2)^2/[(\Nc-1)(\Nc+5)]$ in basis $\mathcal A$ and converges to $16/(5\sqrt5)=1.43$ from above ($2.24$, $1.75$, $1.61$, $1.55$ at $\Nc=3,5,7,9$), whereas in basis $\mathcal B$ it converges to the same limit from below ($0.22$, $0.95$, $1.16$, $1.26$). At $\Nc=3$ the two differ by a factor of ten. We draw two conclusions. The eigenchannel relations are LO predictions of the $1/\Nc$ expansion in the standard evaluation at $\Nc=3$, valid for any function of $G_1\!\cdot\!G_2$, but they are violated by an operator of relative order $1/\Nc^2$ whose matrix elements in the $\Delta$ channels are of order one. The flip relations are shared by the LO and by both truncations, and they are the safest large-$\Nc$ input for the $\Delta$ channels.

\begin{table}[t]
\caption{Diagonal matrix elements at $\Nc=3$ of the leading operator $\Mop$ and of the first correction $\Jop+\Iop$, and their ratio. The nominal suppression is $1/\Nc^2=0.11$.}
\label{tab:size}
\begin{ruledtabular}\small
\begin{tabular}{lccc}
Channel & $\langle\Mop\rangle$ & $\langle\Jop+\Iop\rangle$ & ratio\\
\hline
$\NN$ $(0,1)$, $(1,0)$ & $-25/12$ & $-1/2$ & 0.24\\
$\DD$ $(0,1)$, $(1,0)$ & $55/12$ & $-13/2$ & 1.42\\
$\ND$ $(1,1)$ & $13/4$ & $-5/2$ & 0.77\\
$\ND$ $(2,2)$ & $-3/4$ & $3/2$ & 2.00\\
$\ND$ $(1,2)$, $(2,1)$ & $-17/12$ & $-1/2$ & 0.35\\
$\DD$ $(1,2)$, $(2,1)$ & $11/12$ & $-7/2$ & 3.82\\
$\DD$ $(0,3)$, $(3,0)$ & $-15/4$ & $-3/2$ & 0.40\\
$\DD$ $(2,3)$, $(3,2)$ & $-3/4$ & $3/2$ & 2.00\\
\end{tabular}
\end{ruledtabular}
\end{table}

Beyond LO, by the time-reversal argument the first corrections carry two factors of $J$ or $I$, and the independent bilinear structures are $\Jop$, $\Iop$, $\Xop$ and $\Sop$. We take the general operator of relative order $1/\Nc^2$ to be $f_1(\Mop)\mathcal O f_2(\Mop)$ with $\mathcal O$ one of these four, symmetrized. On the 18-dimensional coupling space the span of the LO together with all such products has dimension 16 at every $\Nc$ from $3$ to $13$. We have not proven that these products exhaust the operators of this order. We have checked numerically, at $\Nc=3$, $5$ and $7$, that two-$J$ or two-$I$ operators with other index contractions and orderings, such as $J_1^iJ_2^jG_1^{ja}G_2^{ia}$, $I_1^aI_2^bG_1^{ib}G_2^{ia}$, $\epsilon^{ijk}\epsilon^{abc}J_1^iG_1^{ja}G_2^{kb}I_2^c$ and the ordered products $J_1^2\,G_1\!\cdot\!G_2$ and $J_1^iI_1^aG_1^{jb}G_2^{ia}G_2^{jb}$, lie inside this span to machine precision, whereas flip-odd operators with four such factors, for instance $(\Jop-\Iop)\Sop$ and $\Jop^2-\Iop^2$, do not. Two relations therefore survive at relative order $1/\Nc^2$ and are first broken at relative order $1/\Nc^4$. They follow from the flip-odd part of the potential, which is $\tfrac12[S(S+1)-I(I+1)]\,g(\Mop)$ with one function $g$ for all sectors, and read
\begin{widetext}
\begin{eqnarray} 
C^{(0,1)}_{\DD\DD}-C^{(0,1)}_{\NN\NN}-\sqrt5\,C^{(0,1)}_{\NN\DD}
&=& C^{(1,0)}_{\DD\DD}-C^{(1,0)}_{\NN\NN}
-\sqrt5\,C^{(1,0)}_{\NN\DD},
\nonumber\\
C^{(1,2)}_{\DD\DD}-C^{(1,2)}_{\ND\ND}+\tfrac{7}{4\sqrt5}\,C^{(1,2)}_{\ND\DD}
&=& C^{(2,1)}_{\DD\DD}-C^{(2,1)}_{\ND\ND}+\tfrac{7}{4\sqrt5}\,C^{(2,1)}_{\ND\DD}.
\label{eq:nnlorel}
\end{eqnarray} 
\end{widetext}
The eigenchannel relations are violated at this order, but equally in the two members of a flip pair: the combination that vanishes at LO takes the same nonzero value in both. The coefficients $\sqrt5$ and $7/(4\sqrt5)$ are the eigenchannel ratios at $\Nc=3$, and for general $\Nc$ they are replaced by the ratios that follow from Tab.~\ref{tab:closed}, with which both relations hold at every $\Nc$ from $3$ to $13$. A minimal scheme keeps one operator per structure,
\begin{equation}
V_{\rm min}=c_0\bone+c_G\Mop+c_J\Jop+c_I\Iop+c_X\Xop+c_S\Sop ,
\label{eq:bil}
\end{equation}
with six parameters. Its flip-odd part is $(c_J-c_I)(\Jop-\Iop)/2$, which is the c-number $\tfrac12(c_J-c_I)\cdot\tfrac12[S(S+1)-I(I+1)]$ times the unit matrix in every sector. The difference between a diagonal coupling and its partner under the spin--isospin flip is therefore the same for every channel $\alpha$ of the pair,
\begin{equation}
C^{(S,I)}_{\alpha\alpha}-C^{(I,S)}_{\alpha\alpha}=\tfrac12(c_J-c_I)\big[S(S+1)-I(I+1)\big],
\label{eq:flipbreak}
\end{equation}
whereas the transition couplings remain flip symmetric. Evaluating the bracket, the flip violations in the pairs $(0,3)$--$(3,0)$, $(2,3)$--$(3,2)$ and $(1,2)$--$(2,1)$ are
\begin{equation}
\begin{aligned}
C^{(0,3)}_{\DD\DD}-C^{(3,0)}_{\DD\DD}&=6\,\big[C^{(0,1)}_{\NN\NN}-C^{(1,0)}_{\NN\NN}\big],\\
C^{(2,3)}_{\DD\DD}-C^{(3,2)}_{\DD\DD}&=3\,\big[C^{(0,1)}_{\NN\NN}-C^{(1,0)}_{\NN\NN}\big],\\
C^{(1,2)}_{\alpha\alpha}-C^{(2,1)}_{\alpha\alpha}&=2\,\big[C^{(0,1)}_{\NN\NN}-C^{(1,0)}_{\NN\NN}\big],
\end{aligned}
\label{eq:ratios}
\end{equation}
with $\alpha=\ND,\DD$ in the last line, where $C^{(0,1)}_{\NN\NN}-C^{(1,0)}_{\NN\NN}=C(\pw{1}{S}{0})-C(\pw{3}{S}{1})$ is the Wigner-symmetry breaking of the $\NN$ channels~\cite{msw1999}. These ratios depend on the truncation of $g$ to a constant and are model statements, which Sec.~\ref{sec:flip} makes explicit. The exact-SU(4) form $V_{\mathcal B}$ is the point $c_J=c_I=c_G$, $c_X=c_S=0$ of this scheme.

\subsection{Boundary condition at $\LH$ and the running of the eigenchannel angle}\label{sec:flow}\label{sec:flipflow}\label{sec:angle}

We now impose the LO structure on the $(0,1)$ sector at the scale $\LH$ and propagate it with the exact flow of Eq.~\eqref{eq:exactsol}. For a $2\times2$ sector every function of $\Mop$ lies in the span of $\bone$ and $\Mop$, so the condition is that $C(\LH)$ has the eigenvectors $n_\pm$ of $\Mop$, with eigen-couplings $v_\pm$. The same holds for the inverse, and the boundary condition reads
\begin{equation}
\begin{aligned}
C^{-1}(\LH)&=\alpha\bone+\beta\Mop,\qquad \frac1{v_\pm}=\alpha+\beta\lambda_\pm,\\
K&=\alpha\bone+\beta\Mop-B(\LH).
\end{aligned}
\label{eq:bc}
\end{equation}
The $\NN$ scattering length fixes one combination of $(\alpha,\beta)$ through $K_{11}-K_{12}^2/K_{22}=\mu_1/(2\pi a)$. We parametrize the remaining freedom by the ratio of eigen-couplings $r_v\equiv v_+/v_-$, with $v_+$ the coupling of the $\DD$-dominant eigenchannel, which is the parameter that the LO leaves free and that no $\NN$ datum can fix by the decoupling of Sec.~\ref{sec:exactflow}. Then $\alpha=c\beta$ with $c=(r_v\lambda_+-\lambda_-)/(1-r_v)$, and the scattering-length condition becomes a quadratic equation for $\beta$ whose physical root is
\begin{equation}
\bh\equiv\frac{\pi^2\beta}{\mu_1\LH}=-\frac{(c+\Mop_{22})\big(1-\frac{\pi}{2a\LH}\big)}{(c+\lambda_+)(c+\lambda_-)}\big[1+\mathcal O(B_2/B_1)\big],
\label{eq:betahat}
\end{equation}
the other root corresponding to $K_{22}\to0$ (App.~\ref{app:family}). The dimensionless $\bh$ is the splitting of the inverse eigen-couplings in units of the $\NN$ scale, $\bh(\lambda_+-\lambda_-)=[(C^{-1})_+-(C^{-1})_-]\pi^2/(\mu_1\LH)$. It is positive when the $\DD$-dominant eigenchannel is the more attractive one and negative otherwise, and it is the only parameter that governs the running of the angle. The transition ratio at the matching scale follows in closed form,
\begin{equation}
\eta(\LH)=\frac{C_{12}}{C_{11}}\bigg|_{\LH}=\frac{2(r_v-1)}{r_v(3-\sqrt5)+3+\sqrt5} ,
\label{eq:rho0}
\end{equation}
independent of $\LH$ and of $a$. It is not the value transmitted to lower cutoffs when $\LH$ is not small compared with $\kDD$, Eq.~\eqref{eq:etaratio}. Tab.~\ref{tab:family} lists four members of the family for $a=a_s$ at $\LH=770$~MeV and, for comparison, at $200$~MeV, together with $\eta$ and $\bar C_{22}$ at $\LH$ and their transmitted values $\eta_K$ and $\bar C_K$. The table shows the point made in Sec.~\ref{sec:exactflow}: for $\LH=770$~MeV the transmitted invariants differ from the values at the matching scale by $8\%$ to $71\%$, because $\LH/\kDD=0.9$, whereas for $\LH=200$~MeV they differ by a few percent.

\begin{table}[t]
\caption{The boundary-condition family of Eq.~\eqref{eq:bc} for the $(0,1)$ sector with $a=a_s$, at $\LH=770$~MeV and at $200$~MeV. Listed are the parameter $\bh$, the transition ratio $\eta$ and the Schur complement $\bar C_{22}$ at $\LH$, their transmitted values $\eta_K$ and $\bar C_K$ of Eq.~\eqref{eq:etaK}, and the eigenchannel angle at $\LL=140$~MeV from the exact flow. The angle at $\LH$ is $\theta_{\mathrm A}=69.1^\circ$ in every row.}
\label{tab:family}
\begin{ruledtabular}\scriptsize\setlength{\tabcolsep}{2pt}
\begin{tabular}{cccccccc}
$\LH$ & $r_v$ & $\bh$ & $\eta(\LH)$ & $\eta_K$ & $\theta(\LL)$ & $\bar C_{22}(\LH)$ & $\bar C_K$\\
(MeV) & & & & & & (fm$^2$) & (fm$^2$)\\
\hline
770 & 1/4 & $-0.315$ & $-0.276$ & $-0.298$ & $72.5^\circ$ & $-0.31$ & $-0.34$\\
770 & 1/2 & $-0.107$ & $-0.178$ & $-0.206$ & $76.8^\circ$ & $-0.59$ & $-0.68$\\
770 & 2   & $+0.066$ & $+0.296$ & $+0.468$ & $35.7^\circ$ & $-1.59$ & $-2.52$\\
770 & 4   & $+0.143$ & $+0.724$ & $+1.240$ & $57.0^\circ$ & $-1.80$ & $-3.08$\\
200 & 1/4 & $-0.324$ & $-0.276$ & $-0.278$ & $70.8^\circ$ & $-1.17$ & $-1.18$\\
200 & 1/2 & $-0.112$ & $-0.178$ & $-0.180$ & $73.6^\circ$ & $-2.19$ & $-2.22$\\
200 & 2   & $+0.067$ & $+0.296$ & $+0.306$ & $56.1^\circ$ & $-6.03$ & $-6.25$\\
200 & 4   & $+0.125$ & $+0.724$ & $+0.758$ & $63.2^\circ$ & $-7.95$ & $-8.33$\\
\end{tabular}
\end{ruledtabular}
\end{table}

Before following the angle, we note which relations the flow cannot touch. The LO flow of a sector depends only on the reduced masses and thresholds of its channels. The two sectors of a flip pair contain the same channels, so that their matrices $B(\Lambda)$ are identical and
\begin{equation}
C^{-1}_{(S,I)}(\Lambda)-C^{-1}_{(I,S)}(\Lambda)=K_{(S,I)}-K_{(I,S)} 
\label{eq:flipinv}
\end{equation}
for all $\Lambda$. If a flip relation holds at one scale it therefore holds at every scale, exactly. The physical violation in the $\NN$ sectors is the $\Lambda$-independent difference of threshold matrices fixed by the data, $(\mu_1/2\pi)(1/a_s-1/a_t)=-3.34\times10^3$~MeV$^2$, whose size relative to the natural scale $\mu_1\Lambda/\pi^2$ is $0.70$, $0.23$ and $0.09$ at $\Lambda=100$, $300$ and $770$~MeV. This is the RG form of the statement that Wigner SU(4) symmetry of the $\NN$ S-waves emerges in the limit of large scattering lengths~\cite{msw1999}: the breaking is a fixed number of order $1/a_s-1/a_t=-44.7$~MeV, and its weight relative to the running coupling falls as $1/\Lambda$. The large-$\Nc$ flip relations can thus be imposed at any scale without creating a scale ambiguity, and the window of subtraction points in which they were found to hold in PDS~\cite{sss2018} is, in this language, the statement that the data-fixed violation is small compared with $\mu_1\mu_{\rm PDS}/(2\pi)$ at large $\mu_{\rm PDS}$.

\begin{figure}[t]
\includegraphics[width=\columnwidth]{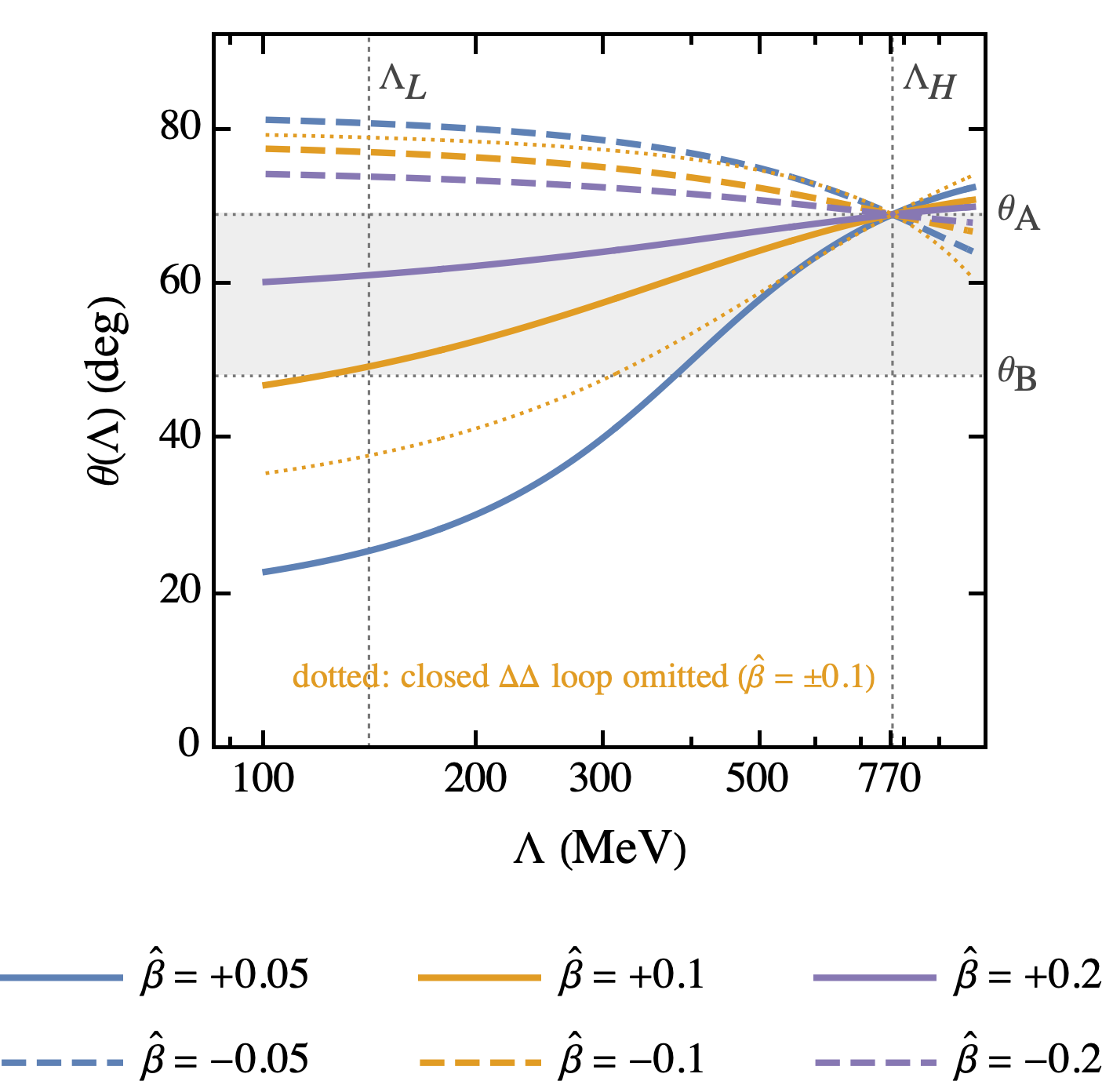}
\caption{Running of the eigenchannel angle of the $\NN$--$\DD$ coupling matrix between the two cutoffs, Eq.~\eqref{eq:theta}, for the boundary condition $\theta(\LH)=\theta_{\mathrm A}$ at $\LH=770$~MeV and six values of $\bh$, positive values as solid and negative values as dashed lines, on a logarithmic cutoff axis. The thin dotted lines show $\bh=\pm0.1$ with the closed $\DD$ loop omitted. The band spans the values of the angle in the two two-parameter forms of Sec.~\ref{sec:LOrel}, from the exact-SU(4) value $\theta_{\mathrm B}$ to the Hartree value $\theta_{\mathrm A}$, and the vertical dotted lines mark $\LL$ and $\LH$.}
\label{fig:univ}
\end{figure}

\begin{figure*}[t]
\includegraphics[width=\textwidth]{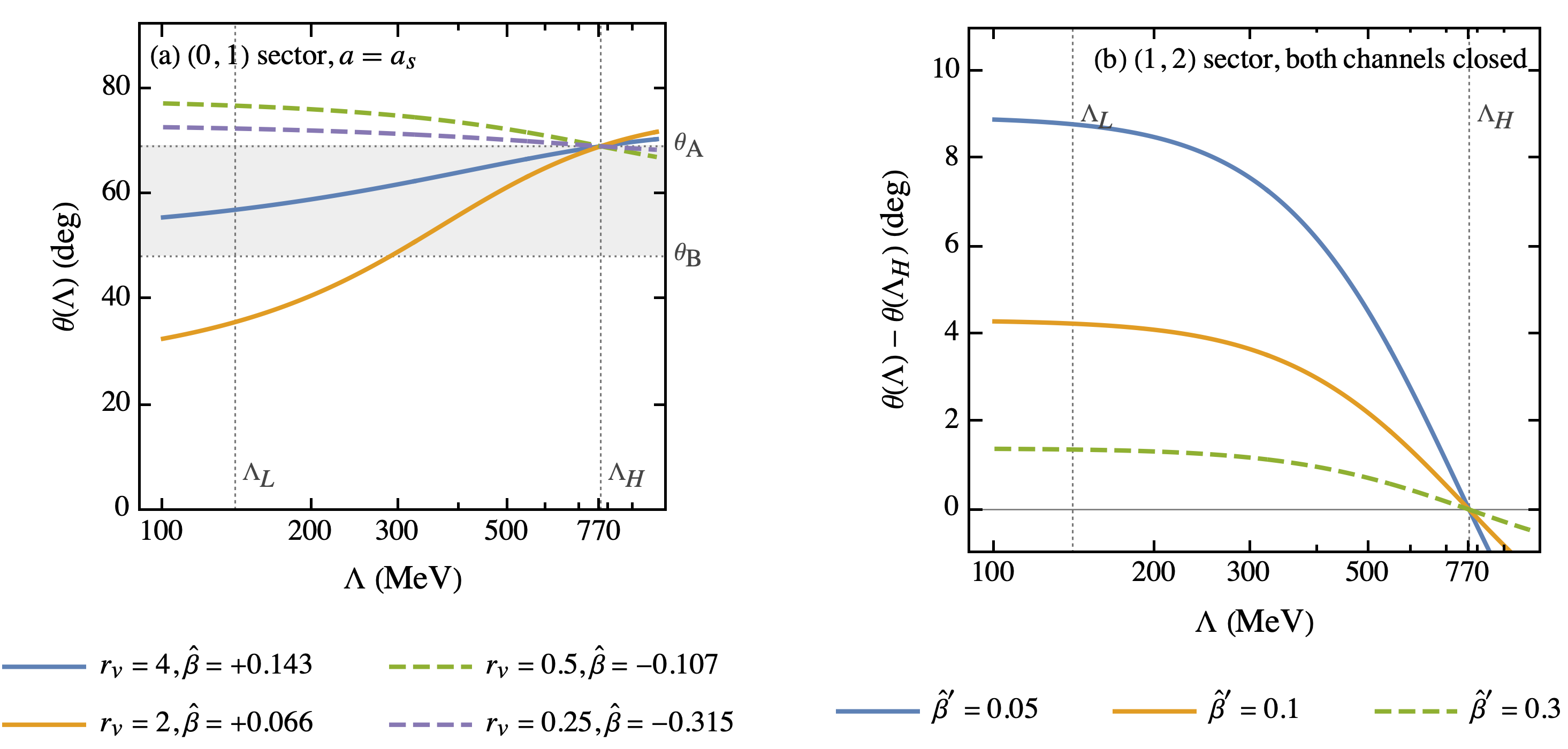}
\caption{(a) Exact running of the eigenchannel angle for the four members of the boundary-condition family of Tab.~\ref{tab:family} with $\LH=770$~MeV and $a=a_s$, from the exact flow of Eq.~\eqref{eq:exactsol}. Solid lines have $\bh>0$ and dashed lines $\bh<0$, and the band and the horizontal dotted lines mark $\theta_{\mathrm B}$ and $\theta_{\mathrm A}$ as in Fig.~\ref{fig:univ}. (b) Change of the angle $\theta(\Lambda)-\theta(\LH)$ in the closed $\ND$--$\DD$ sector $(1,2)$ for three values of $\bh'$, with the boundary condition of App.~\ref{app:family} and $\alpha'=-\mu_{\ND}\LH/\pi^2$. The vertical dotted lines mark $\LL$ and $\LH$.}
\label{fig:exact}
\end{figure*}

The eigenchannel relations behave differently. With $K$ from Eq.~\eqref{eq:bc}, $C^{-1}(\Lambda)=\alpha\bone+\beta\Mop+B(\Lambda)-B(\LH)$, and the angle of its eigenvector connected to $n_+$ obeys
\begin{equation}
\begin{aligned}
\tan2\theta(\Lambda)&=\frac{2\Mop_{12}}{(\Mop_{11}-\Mop_{22})-\Delta_H(\Lambda)/\bh},\\
\Delta_H(\Lambda)&\equiv\frac{\Lambda-\LH}{\LH}-\frac{\mu_2}{\mu_1}\frac{F(\Lambda)-F(\LH)}{\LH},
\end{aligned}
\label{eq:theta}
\end{equation}
with $F(\Lambda)=\Lambda-\kDD\arctan(\Lambda/\kDD)$, $\Mop_{11}-\Mop_{22}=-\tfrac{20}3$ and $2\Mop_{12}=\tfrac{8\sqrt5}3$. Equation~\eqref{eq:theta} is an exact identity for the eigenvectors of $C^{-1}(\Lambda)$, and hence of $C(\Lambda)$, for a given $\beta$. The first term of $\Delta_H$ is the running of the open $\NN$ loop, the second that of the closed $\DD$ loop. Fig.~\ref{fig:univ} shows $\theta(\Lambda)$ between the two cutoffs for the boundary condition $\theta(\LH)=\theta_{\mathrm A}$ and six values of $\bh$, and Fig.~\ref{fig:exact}(a) the exact running for the four members of Tab.~\ref{tab:family}. The physical origin of the running is simple. The inverse coupling of the open channel runs linearly, $(C^{-1})_{11}=K_{11}-\mu_1\Lambda/\pi^2$, whereas the $\DD$ entries are nearly frozen, and a matrix whose diagonal entries drift apart at a fixed off-diagonal element rotates its eigenvectors toward the channel axes. The size of the running is controlled by $\bh$, which the $\NN$ data do not fix. Equation~\eqref{eq:theta} shows that the angle at $\LL$ is a function of $\Delta_H(\LL)/\bh$ alone, with $\Delta_H(\LL)=-0.57$ for $\LH=770$ and $\LL=140$~MeV. For $|\bh|\gg|\Delta_H|$ the eigen-couplings are strongly split, the denominator is dominated by $\Mop_{11}-\Mop_{22}$, and the angle stays close to $\theta_{\mathrm A}$. For $|\bh|\lesssim|\Delta_H|$ the running term dominates and the angle moves toward the channel axes, $\theta\to0$ for $\bh>0$ and $\theta\to90^\circ$ for $\bh<0$ as $|\bh|\to0$. The numbers in Tab.~\ref{tab:family} and Figs.~\ref{fig:univ} and \ref{fig:exact}(a) illustrate this: between $\LH$ and $\LL$ the angle moves from $69.1^\circ$ to $49.4^\circ$, $61.2^\circ$ and $77.2^\circ$ for $\bh=0.1$, $0.2$ and $-0.1$, and to $36^\circ$, $57^\circ$, $77^\circ$ and $73^\circ$ for the four members of the family. The eigenchannel relation $(C_{\DD}-C_{\NN})/\Ctr=\sqrt5$ therefore holds at the scale at which it is imposed and at no other, and without an assumption on $\bh$ it cannot be imposed at a definite scale. For the values of $\bh$ in Tab.~\ref{tab:family} the scale ambiguity between the two cutoffs is of the same size as the $21^\circ$ spread between $\theta_{\mathrm A}$ and $\theta_{\mathrm B}$. In the closed $\ND$--$\DD$ sector $(1,2)$ we follow the eigenvector connected to $n_+$ of $\Mop_{(1,2)}$, whose angle is $34.3^\circ+90^\circ=124.3^\circ$ in the convention of Eq.~\eqref{eq:thetaA}. For the boundary condition $C^{-1}(\LH)=\alpha'\bone+\beta'\Mop_{(1,2)}$ with $\alpha'=-\mu_{\ND}\LH/\pi^2$ and $\beta'=\bh'\mu_{\ND}\LH/\pi^2$, this angle moves by $4.3^\circ$ for $\bh'=0.1$ and by $8.8^\circ$ for $\bh'=0.05$ over the same window, almost all of it above $300$~MeV, Fig.~\ref{fig:exact}(b). Relations among closed channels only are scale independent in the decoupled regime, where all their couplings run by the same $\Lambda^3/\kappa^2$ terms, but not between $770$ and $140$~MeV, where neither closed loop is small: the accidental relation $C^{(2,2)}_{\ND}=C^{(2,3)}_{\DD}$ imposed at $\LH$ with couplings of natural size is violated by $14\%$ at $\LL$, because the two channels have different closed-channel momenta, $\kND=559$ and $\kDD=850$~MeV. These violations are quantitative statements, recorded in Tab.~\ref{tab:relations}, and for moderate $|\bh|$ they are comparable to the running in the $(0,1)$ sector.

\subsection{What controls the running: the closed-channel momentum and the large-$\Nc$ counting}\label{sec:degenerate}\label{sec:inherited}

The running of the angle in Eq.~\eqref{eq:theta} is governed by the difference between the closed $\DD$ loop and the open $\NN$ loop at the $\NN$ threshold,
\begin{equation}
\begin{aligned}
\Delta B(\Lambda)&\equiv B_2(0;\Lambda)-B_1(0;\Lambda)=\frac{\mu_1\Lambda-\mu_2F(\Lambda)}{\pi^2},\\
\tan2\theta(\Lambda)&=\frac{2\beta\Mop_{12}}{\beta(\Mop_{11}-\Mop_{22})-[\Delta B(\Lambda)-\Delta B(\LH)]}.
\end{aligned}
\label{eq:D}
\end{equation}
Two parameters enter $\Delta B$, the ratio of reduced masses $\mu_2/\mu_1=1+\dM/\MN$ and the closed-channel momentum $\kDD=\sqrt{2\MD\dM}$ in $F$. The loop difference vanishes identically only when both $\mu_2=\mu_1$ and $\kDD=0$, that is, in the formal degenerate limit $\dM\to0$ at fixed baryon mass, in which the two loops coincide, $B(\Lambda)\propto\bone$, and every relation that is linear in $C^{-1}$ with $\Lambda$-independent coefficients, including $C^{-1}\in\mathrm{span}\{\bone,\Mop\}$, is an invariant of the flow. At the physical splitting $\Delta B(\Lambda)$ is $0.99$, $0.95$ and $0.75$ times $\mu_1\Lambda/\pi^2$ at $\Lambda=100$, $300$ and $770$~MeV, $\Delta B(\LH)-\Delta B(\LL)=2.1\times10^4$~MeV$^2$ is comparable to $\beta(\Mop_{11}-\Mop_{22})=-2.4\times10^4\,(\bh/0.1)$~MeV$^2$, and the running is of order $10^\circ$ to $30^\circ$.

\begin{figure*}[t]
\includegraphics[width=\textwidth]{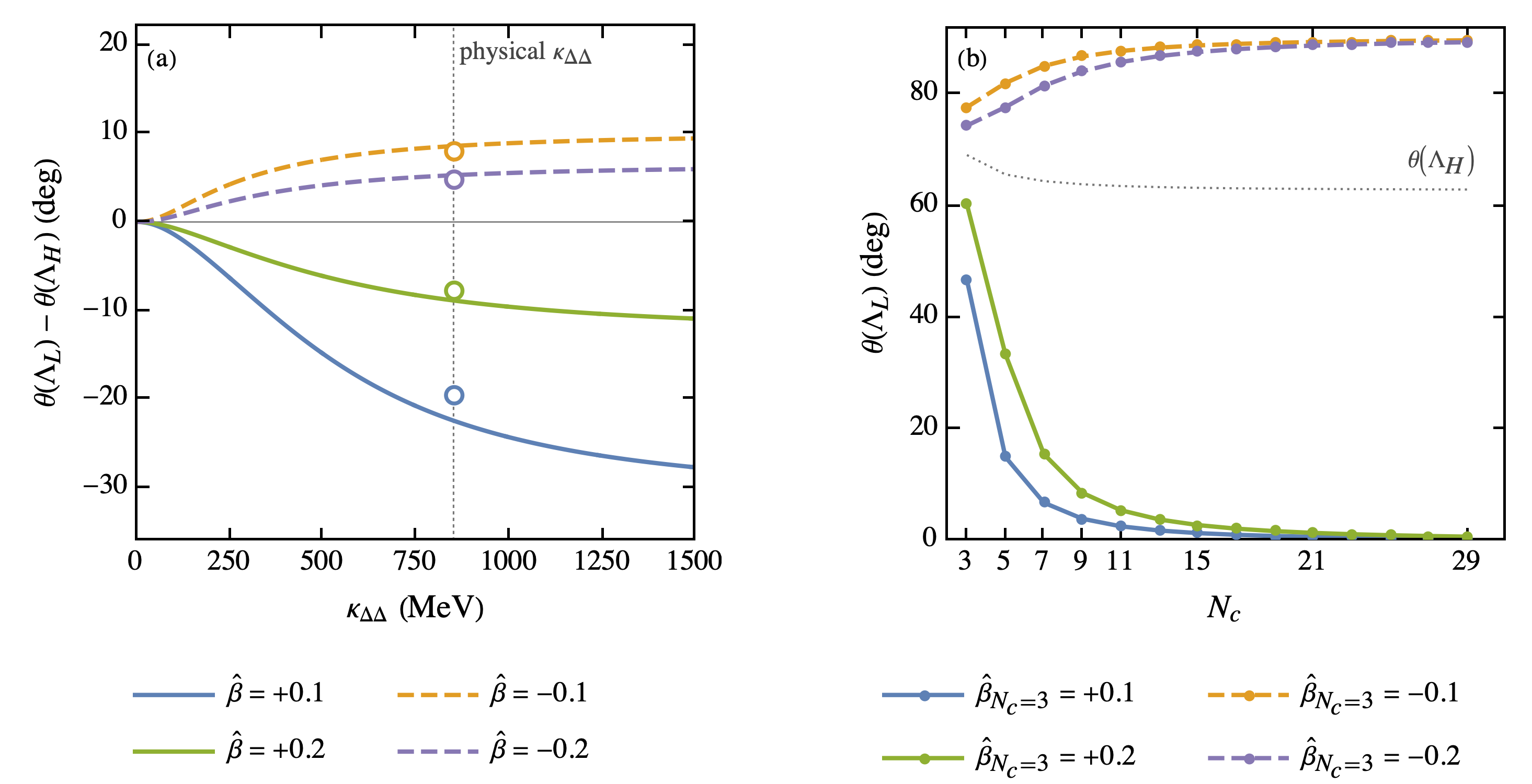}
\caption{(a) Change of the eigenchannel angle between $\LH=770$~MeV and $\LL=140$~MeV as a function of the closed-channel momentum $\kDD$ at equal reduced masses, $\mu_2=\mu_1$, for four values of $\bh$, positive values as solid and negative values as dashed lines. The circles mark the physical point, $\mu_2=\MD/2$ and $\kDD=850$~MeV, and the vertical dotted line the physical $\kDD$. The running vanishes only as $\kDD\to0$. (b) The angle at $\LL$ for odd $\Nc$ from $3$ to $29$, for the general-$\Nc$ matrices of Tab.~\ref{tab:closed}, with $\mu_2=\mu_1$, $\kDD$ at its physical value, and the splitting of the inverse eigen-couplings in units of the loop scaled as $\Nc^{-2}$ from its value $\bh_{\Nc=3}$ at $\Nc=3$, as the Hartree counting requires. The dotted line is the angle $\theta(\LH)$ imposed at the high cutoff.}
\label{fig:deg}
\end{figure*}

The degenerate limit is not the large-$\Nc$ limit. In the $1/\Nc$ expansion the baryon masses grow as $\MN,\MD\sim\Nc\Lambda_{\rm QCD}$ while the splitting falls as $\dM\sim\Lambda_{\rm QCD}/\Nc$~\cite{djm1994}, so that
\begin{equation}
\frac{\mu_2}{\mu_1}=1+\mathcal O\Big(\frac1{\Nc^2}\Big),\qquad
\kDD^2=2\MD\dM=\mathcal O(\Nc^0).
\label{eq:kappaNc}
\end{equation}
As $\Nc\to\infty$ the reduced masses become equal, but the closed-channel momentum stays at its physical size, and with it the loop difference, $\Delta B(\Lambda)\to\mu_1[\kDD\arctan(\Lambda/\kDD)]/\pi^2$, which is of order $\Nc$ like the loops themselves. Fig.~\ref{fig:deg}(a) shows the change of the angle between the two cutoffs as a function of $\kDD$ at $\mu_2=\mu_1$, with the physical point, $\mu_2\ne\mu_1$ and $\kDD=850$~MeV, marked. The running at $\mu_2=\mu_1$ and physical $\kDD$ is $-22.4^\circ$, $-8.9^\circ$ and $+8.6^\circ$ for $\bh=0.1$, $0.2$ and $-0.1$, against $-19.7^\circ$, $-7.9^\circ$ and $+8.1^\circ$ at the physical point: the reduced-mass difference accounts for a small part of the running, and the threshold momentum for the rest. The running goes to zero only as $\kDD\to0$, and at $\kDD=850$~MeV it has not saturated, since $F(\Lambda)$ still depends on $\kDD$ at $\Lambda\sim\kDD$. Along the trajectory $\MN\propto\Nc$, $\dM\propto1/\Nc$, normalized to the physical values at $\Nc=3$, the closed-channel momentum tends to $\sqrt{2\MN\dM}\,|_{\Nc=3}=742$~MeV and $\Delta_H(\LL)$ to $-0.60$, and at fixed $\bh$ the change of the angle grows slightly, to $-20.7^\circ$, $-8.2^\circ$ and $+8.3^\circ$ for $\bh=0.1$, $0.2$ and $-0.1$. The $1/\Nc$ expansion therefore does not remove the scale dependence of the eigenchannel relations.

The counting makes the statement sharper. In the Hartree form of Eq.~\eqref{eq:hartree} the couplings are of order $\Nc$, so that $C^{-1}$, its eigenvalues and the splitting $\beta(\lambda_+-\lambda_-)$ of the inverse eigen-couplings are of order $1/\Nc$, whereas the loops $\mu\Lambda/\pi^2$ and their difference are of order $\Nc$. In Eq.~\eqref{eq:D} the term $\beta(\Mop_{11}-\Mop_{22})$ is thus suppressed by $\Nc^{-2}$ relative to $\Delta B(\Lambda)-\Delta B(\LH)$, and in the limit the angle is rotated all the way to the channel axes, $\theta(\LL)\to0$ or $90^\circ$ according to the sign of $\bh$. Fig.~\ref{fig:deg}(b) shows this for the general-$\Nc$ matrices of Tab.~\ref{tab:closed} with the splitting of the inverse eigen-couplings scaled as $\Nc^{-2}$ from its $\Nc=3$ value: for $|\bh_{\Nc=3}|=0.1$ the angle at $\LL$ moves from $47^\circ$ and $78^\circ$ at $\Nc=3$ to $7^\circ$ and $85^\circ$ at $\Nc=7$, and for $|\bh_{\Nc=3}|=0.2$ from $60^\circ$ and $74^\circ$ to $16^\circ$ and $82^\circ$. A consistent large-$\Nc$ counting maximizes the violation of the eigenchannel relations by the flow rather than removing it. This is the two-baryon form of the observation of Refs.~\cite{bcg2002,bc2002} that the large-$\Nc$ relations constrain the potential and not the amplitude, since with $\MN\sim\Nc$ and $C\sim\Nc$ the iterated loop is enhanced by $\Nc^2$ relative to $C$. Here the enhancement is the linear divergence $\mu_1\Lambda/\pi^2$ that runs with the cutoff and differs between the open and the closed channel.

The eleven LO relations therefore fall into three classes under the flow, as recorded in Tab.~\ref{tab:relations}. The eight flip relations are exact invariants at every $\Nc$ and every $\kDD$, because the two sectors of a flip pair have identical loops. The relations among closed channels only are invariant in the decoupled regime and violated at the $10\%$ level between $770$ and $140$~MeV. The eigenchannel relation of the $(0,1)$ sector, which connects the open $\NN$ channel to the closed $\DD$ channel, runs by an amount controlled by $\kDD$ and by $\bh$, which is of order one at the physical point and grows with $\Nc$ in the Hartree counting. It can be imposed at one scale only, and the $1/\Nc$ expansion does not say at which. The comparison with Ref.~\cite{rss2025} is then the following. There $\MN=\MD$ is set, so that $\kDD=0$ and the loops are identical, the LO structure is an invariant of the flow, and the requirement of scale independence of all channel amplitudes together with unitarity drives the theory to the rank-two fixed point of Sec.~\ref{sec:fixedpoints}, at which $C\propto\bone$. At finite $\kDD$ the $\NN$ channel flows instead to the single-channel fixed points of Ref.~\cite{bmr1999} with the $\DD$ channel as a spectator, and the two descriptions differ at $\Lambda\lesssim\kDD$ by terms that are of order one at the physical point.

\begin{figure*}[t]
\includegraphics[width=\textwidth]{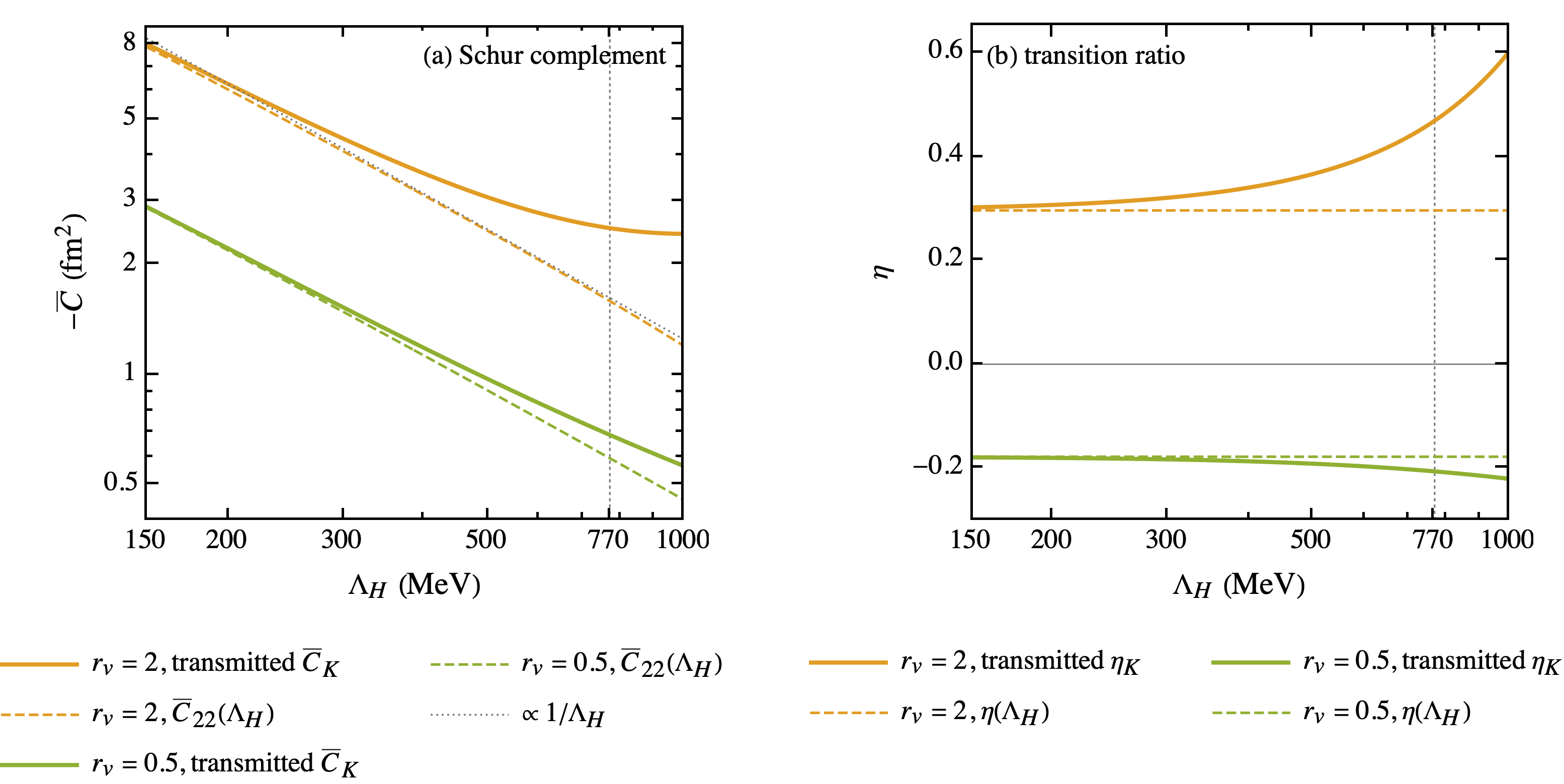}
\caption{Invariants of the $\Delta$ channels when the LO structure $\theta=\theta_{\mathrm A}$ is imposed at the matching scale $\LH$ with fixed eigen-coupling ratio $r_v$ and $a=a_s$, for $r_v=2$ (upper curves) and $r_v=0.5$ (lower curves). (a) Schur complement, Eq.~\eqref{eq:inv}. Solid lines show the transmitted value $\bar C_K$ of Eq.~\eqref{eq:etaK}, dashed lines the value $\bar C_{22}(\LH)$ at the matching scale, and the dotted line a $1/\LH$ dependence normalized to the $r_v=2$ curve at $200$~MeV. (b) Transition ratio. Solid lines show the transmitted $\eta_K$, dashed lines $\eta(\LH)$ of Eq.~\eqref{eq:rho0}, which does not depend on $\LH$. The vertical dotted line is $\LH=770$~MeV.}
\label{fig:inv}
\end{figure*}

Finally, what the LO boundary condition at $\LH$ transmits to the pionless theory is the threshold matrix $K$, equivalently $(\eta_K,\bar C_K)$, together with the $\NN$ datum. Equation~\eqref{eq:rho0} gives $\eta$ at $\LH$, which depends on $r_v$ alone, and Eq.~\eqref{eq:etaratio} the transmitted values. Fig.~\ref{fig:inv} shows them as functions of the matching scale. With $v_-\simeq-\pi^2/(\mu_1\LH)$ fixed by the near-unitary $\NN$ datum, $v_+=r_vv_-$ scales as $1/\LH$ and so does $\bar C_{22}(\LH)$, whereas the transmitted $\bar C_K$ and $\eta_K$ depend in addition on $\LH$ through the closed loop, which is visible above $300$~MeV. The LO large-$\Nc$ structure imposed at $\LH$ with eigen-coupling ratio $r_v$ thus fixes $\eta$ at $\LH$, and the transmitted invariants depend on $\LH$. Neither the size of the $\DD$ coupling ($r_v$) nor the scale at which the structure is imposed ($\LH$) is fixed by the expansion, and the $\NN$ observables are insensitive to both at the level of $\Lambda^2/\kDD^2$.


%% ================================================================
\section{Matching between the two cutoffs and flip-breaking relations}\label{sec:twocut}
%% ================================================================

\subsection{The two cutoffs and what the flow transmits}\label{sec:window}\label{sec:transmit}\label{sec:matching}

We now place the results of Secs.~\ref{sec:decoupling} and \ref{sec:largeN} in the setting that motivates them. At the high cutoff $\LH\lesssim m_V$ the contact couplings of the $\Delta$-full chiral theory are the natural target of resonance saturation, and at the low cutoff $\LL=m_\pi$ they are the couplings of pionless EFT. In the contact sector the two sets are connected by the exact flow of Eq.~\eqref{eq:exactsol},
\begin{equation}
C^{-1}(\LL)=C^{-1}(\LH)+B(0;\LL)-B(0;\LH),
\label{eq:match}
\end{equation}
which holds for any value of $\Lambda/\kDD$ and replaces a matching of amplitudes at one scale. Four scales lie inside the window and decouple one by one as the cutoff passes below them: $\kDD$, at which the ratio of the closed to the open loop of Sec.~\ref{sec:bubble} is $0.25$ for $\LH=770$~MeV and $0.01$ for $\LL=140$~MeV, $\kND$, the splitting $\dM$, and the pion mass. The first three are treated exactly by Eq.~\eqref{eq:match}. The pion is not, and its status is as follows. The S-wave central part of one-pion exchange between nucleons is, in the sign convention of App.~\ref{app:conv} in which a positive potential is repulsive,
\begin{equation}
V_\pi^{\rm c}(q)=-\frac{g_A^2}{4f_\pi^2}\,\bm\tau_1\!\cdot\!\bm\tau_2\,\frac13\,\bm\sigma_1\!\cdot\!\bm\sigma_2\,\frac{q^2}{q^2+m_\pi^2},
\label{eq:ope}
\end{equation}
with $g_A^2/(4f_\pi^2)=4.7\times10^{-5}$~MeV$^{-2}$ for $g_A=1.27$ and $f_\pi=92.4$~MeV. For $q\ll m_\pi$ it is of order $q^2/m_\pi^2$, because the zero-momentum contact piece of the pion is cancelled by the Yukawa at $q=0$, so that the pion decouples from the LO S-wave flow with a relative correction of order $(\Lambda/m_\pi)^2$, in the same way as the closed $\DD$ channel decouples with $\Lambda^2/\kDD^2$. With $\LL=m_\pi$ this correction factor is one, and Eq.~\eqref{eq:match} at $\LL$ is a statement about the contact couplings of a theory in which the pion is still present. Inside the window the pion is not a perturbation on the S-wave contact above a few hundred MeV: taking $q\sim\Lambda$ as a scale estimate and $\langle\bm\tau_1\!\cdot\!\bm\tau_2\,\bm\sigma_1\!\cdot\!\bm\sigma_2\rangle=-3$ in $\pw{1}{S}{0}$, the ratio $|V_\pi^{\rm c}(\Lambda)|/|C_{\pw{1}{S}{0}}(\Lambda)|$ is $0.17$, $0.58$ and $1.7$ at $\Lambda=140$, $300$ and $770$~MeV, consistent with the conclusion of Ref.~\cite{ks1999} that one-pion exchange is perturbative in $\pw{1}{S}{0}$ away from the amplitude pole and that the failure of the expansion above the pion mass can be due to short-range physics. In the $\pw{3}{S}{1}$ channel the tensor force is not perturbative in this window~\cite{ntvk2005}. The contact-sector matching of this work is therefore exact as a statement about the contact theory, and the running of $10^\circ$ to $30^\circ$ found in Sec.~\ref{sec:largeN} is the running of that theory. It becomes a statement about nature once the pion is included in the flow, with the distorted-wave RG of Refs.~\cite{bb2003,birse2006} and the $\pi N\Delta$ coupling, which is left to future work. The saturation picture of Ref.~\cite{egme2002} refers to the LECs of a theory that keeps pion exchange explicit, so that identifying those LECs with the couplings of the contact theory at $\LH$ mixes two different theories, which is a further reason to read Sec.~\ref{sec:saturation} as a structural statement only.

What Eq.~\eqref{eq:match} transmits unchanged from $\LH$ to $\LL$ is summarized in Tab.~\ref{tab:relations} and Sec.~\ref{sec:exactflow}. It transmits the eight flip relations, because the two sectors of a flip pair have the same $B(\Lambda)$, which is a kinematic identity that holds for any energy-independent potential. It transmits the elements $K_{12}$ and $K_{22}$ of the threshold matrix of each $\NN$--$\DD$ sector exactly, and these reduce to the transition ratio $\eta$ and the Schur complement $\bar C_{22}$ only in the decoupled regime $\Lambda\ll\kDD$. It transmits the boundary data of the six closed sectors up to corrections of relative order $\Lambda^2/\kappa^2$. It does not transmit the eigenchannel relations, nor any relation among couplings of channels with different loops: between $770$ and $140$~MeV the eigenchannel angle of the $(0,1)$ sector moves from $69.1^\circ$ to $49.4^\circ$, $61.2^\circ$ and $77.2^\circ$ for $\bh=0.1$, $0.2$ and $-0.1$ (Sec.~\ref{sec:degenerate}). The fine tuning of the $\NN$ scattering lengths quoted in the Introduction follows from Eq.~\eqref{eq:match} in one channel, $1/C(\LL)-1/C(\LH)=(\mu_1/\pi^2)(\LH-\LL)=3.0\times10^4$~MeV$^2$, against $(\mu_1/2\pi)(1/a)=-621$ and $2720$~MeV$^2$ in $\pw{1}{S}{0}$ and $\pw{3}{S}{1}$. A boundary value at $\LH$ would have to be accurate to 2\% and 8\% to fix $a_s$ and $a_t$, which is why this program predicts relations among couplings and the couplings of the $\Delta$ channels given the $\NN$ input, and not the $\NN$ scattering lengths.

One property of the pion can be stated without solving the flow. Between S-wave states only the central part of the exchange $(\bm X_1\!\cdot\!\bm q)(\bm Y_2\!\cdot\!\bm q)$ contributes, since the angular projection replaces $q^iq^j$ by $q^2\delta^{ij}/3$ plus a rank-two tensor that connects $L=0$ only to $L=2$. In every $N$--$\Delta$ channel the central part is the product of a spin scalar and an isospin scalar built from the same matrices, $(\bm\sigma,\bm\tau)$, $(\bm S^\dagger,\bm T^\dagger)$ or $(\bm\Sigma,\bm\Theta)$, with coupling constants that do not depend on $S$ or $I$~\cite{sk2020}. By Tab.~\ref{tab:factors} its matrix in the sector $(S,I)$ equals its matrix in $(I,S)$. The two sectors of a flip pair then have the same loops dressed by the central one-pion exchange, and Eq.~\eqref{eq:flipinv} holds exactly with the free loops replaced by the dressed ones. The flip relations are therefore transmitted exactly also when the central one-pion exchange is iterated in the S-waves. The tensor part breaks this. It couples the S-waves to D waves whose content differs between the two members of a flip pair, for example $\pw{1}{S}{0}$ has no D-wave partner whereas $\pw{3}{S}{1}$ couples to $\pw{3}{D}{1}$, and the violation of the flip relations by the pion is thus of second order in the tensor operator.

The same quantities are what the $\Delta$-full chiral theory inherits from the contact sector. In the $\delta$ counting of Pascalutsa and Phillips~\cite{pp2003,pvy2007}, with $\delta\equiv\dM/\Lambda_\chi$, $\Lambda_\chi\simeq1$~GeV and $m_\pi/\Lambda_\chi\sim\delta^2$, the $\Delta$ propagator at $p\sim m_\pi$ counts as $1/\dM$ rather than $1/p$, and the matching of the $\Delta$-full to the $\Delta$-less theory in the one-baryon sector is a decoupling hierarchy in which the $\Delta$-pole contributions to the $\pi N$ constants scale as $c_i^\Delta\sim h_A^2/\dM$, $d_i^\Delta\sim h_A^2/\dM^2$ and $e_i^\Delta\sim h_A^2/\dM^3$. In the two-baryon sector the corresponding threshold is not $\dM$ but $\kDD=\sqrt{2\MD\dM}$, which in this counting is $\kDD/\Lambda_\chi=\sqrt{2(\MD/\Lambda_\chi)\delta}\simeq0.85$, that is, $\kDD\propto\delta^{1/2}\Lambda_\chi$, a hard scale. The external $\Delta$ channels in S-waves are thus integrated out at the same level as $\Lambda_\chi$ physics, and their effect is Eq.~\eqref{eq:match} rather than an expansion in $1/\dM$. The $\Delta$-full theory with cutoffs of $450$--$700$~MeV~\cite{sk2020} inherits $K$ and the $\NN$ datum in each $\NN$--$\DD$ sector, and the couplings of the closed sectors, with the large-$\Nc$ structure entering only through what is imposed on them at $\LH$.

\subsection{The saturated boundary condition at order $q^0$}\label{sec:saturation}

We next ask what structure resonance saturation gives to the sector couplings at $\LH$, assuming that the contact couplings at the high cutoff are the low-momentum limit of scalar ($\sigma$), $\omega$ and $\rho$ exchange with the baryon vertices at LO in $q/M$, and that the couplings to the $\Delta$ are related to those of the nucleon by SU(6), or equivalently at LO in $1/\Nc$, where the $\omega$ couples to baryon number and the $\rho$ (vector coupling) to the isospin generator. At LO in $q/M$ the vertices of scalar and vector exchange are spin independent and the isovector vertex is the isospin generator $I^a$ of the baryon, so that (App.~\ref{app:obe})
\begin{equation}
\begin{aligned}
V^{(0)}(q)&=-\frac{g_\sigma^2}{q^2+m_\sigma^2}+\frac{g_\omega^2}{q^2+m_\omega^2}+\frac{g_\rho^2}{q^2+m_\rho^2}\,4I_1\!\cdot\!I_2\\
&\xrightarrow{\ q\ll m\ }\ C_\sigma+C_\omega+C_\rho\,\langle4I_1\!\cdot\!I_2\rangle_I,
\end{aligned}
\label{eq:sat}
\end{equation}
with $C_\sigma=-g_\sigma^2/m_\sigma^2$, $C_\omega=g_\omega^2/m_\omega^2$ and $C_\rho=g_\rho^2/m_\rho^2$, where $4I_1\!\cdot\!I_2$ is $\bm\tau_1\!\cdot\!\bm\tau_2$, $\bm\tau_1\!\cdot\!\bm\Theta_2$ and $\bm\Theta_1\!\cdot\!\bm\Theta_2$ for $\NN$, $\ND$ and $\DD$, with the eigenvalues of Tab.~\ref{tab:factors}. The spin-dependent pieces of vector exchange are of order $q^2/M^2$ and, for the $\rho$, carry $(1+\kappa_\rho)^2$ with a large tensor-to-vector ratio $\kappa_\rho$, so that at $q\sim\LH$ they are not small, but they are momentum dependent and belong to the $C^{(2)}$ terms of Eq.~\eqref{eq:ctNLO}, not to the $C^{(0)}$ terms. Within these assumptions, Eq.~\eqref{eq:sat} has a consequence that follows from group theory alone. The operators $\bone$ and $I^a$ have no matrix elements between the nucleon and the $\Delta$, because $I^a$ is a generator of the isospin multiplet. The operators that connect them are the spin--isospin generators $G^{ia}$, which enter one-boson exchange only through the axial coupling of the pion, $G^{ia}q^i$, and through the tensor coupling of the $\rho$, both of order $q$. Hence, at order $q^0$ and LO in $q/M$,
\begin{equation}
C^{(S,I)}_{\NN\DD}(\LH)=C^{(S,I)}_{\ND\DD}(\LH)=0\,,
\label{eq:notr}
\end{equation}
in every sector and the diagonal couplings $C^{(S,I)}_{\alpha\alpha}(\LH)=C_\sigma+C_\omega+C_\rho\langle4I_1\!\cdot\!I_2\rangle_I$ are independent of $S$. This structure is an artifact of the $q^0$ truncation. In the standard large-$\Nc$ counting of the meson--baryon couplings~\cite{ks1996}, the vector coupling of the $\rho$ to the isospin current is suppressed by $1/\Nc$ relative to the pion coupling, $g_A\sim\Nc$, and to the tensor coupling of the $\rho$, whose ratio $\kappa_\rho$ to the vector coupling is of order $\Nc$ because the isovector magnetic moment is. The exchanges that are leading in $1/\Nc$ are precisely those that couple through $G^{ia}$ and carry a power of $q$, and at $q\sim\LH\gg m_\pi$ one-pion exchange acts as a contact interaction of the $G_1\!\cdot\!G_2$ type that connects $N$ and $\Delta$. Keeping the $q^0$ terms at $q\sim m_V$ therefore retains the $\Nc$-subleading exchanges and drops the $\Nc$-leading ones, and the absence of a transition coupling in Eq.~\eqref{eq:notr} says nothing about the physical transition coupling at $\LH$. The comparison also fails quantitatively. The flip-odd part of Eq.~\eqref{eq:sat} has the pattern of Eq.~\eqref{eq:ratios}, $C^{(0,1)}_{\NN\NN}-C^{(1,0)}_{\NN\NN}=4C_\rho$, $C^{(1,2)}_{\alpha\alpha}-C^{(2,1)}_{\alpha\alpha}=8C_\rho$ and $C^{(0,3)}_{\DD\DD}-C^{(3,0)}_{\DD\DD}=24C_\rho$, in the ratios $1:2:6$ of the minimal scheme, but with $g_\rho^2/4\pi$ between $0.5$ and $1$, the range of one-boson-exchange models~\cite{mhe1987}, $C_\rho=g_\rho^2/m_\rho^2=1$--$2\times10^{-5}$~MeV$^{-2}$, whereas the value implied by the LO couplings of Eq.~\eqref{eq:CLambda} at $\LH=770$~MeV, $(C_s-C_t)/4=6.6\times10^{-7}$~MeV$^{-2}$, is smaller by a factor of $16$ to $32$. The near-unitary LO couplings $C_s$ and $C_t$ differ by $10\%$ at this cutoff, and their difference is set by $1/a_s-1/a_t$ and not by a tree-level exchange. We conclude that a saturated boundary condition at order $q^0$ with the lightest exchanges has neither the leading large-$\Nc$ structure nor the size of the LO couplings of the contact theory at $\LH$, and that a quantitative saturation analysis requires the full one-boson-exchange potential with its momentum dependence at $q\sim\LH$, projected on the $(S,I)$ sectors, which is a different calculation. The structural statement that survives is Eq.~\eqref{eq:notr} itself: the contact flow of Eq.~\eqref{eq:match} keeps $C^{-1}$ diagonal if it is diagonal at one scale, so that the transition couplings of the $\Delta$ channels at any cutoff below $\LH$ have to be generated by the momentum-dependent exchanges inside the window, of which one-pion exchange is the part the EFT controls.

\subsection{The $\DD$ flip relation and its first correction}\label{sec:flip}

The flip relations are the large-$\Nc$ statements that the flow transmits exactly, and their first corrections are therefore the natural place to look for a prediction among scale-independent quantities. The cleanest flip pair is $(S,I)=(3,0)$ and $(0,3)$: both are single $\DD$ channels with identical loops, so that the relation reads directly on the scattering lengths, and both are exotic, with no $\NN$ admixture. The $(3,0)$ channel also contains the $d^*(2380)$ resonance~\cite{adlarson2011}, about $80$~MeV below the $\DD$ threshold, whose binding momentum of about $320$~MeV lies above $m_\pi$ and outside the pionless theory, and no input from it or from any specific dibaryon enters the relation. At LO,
\begin{equation}
\frac{1}{a^{(0,3)}_{\DD}}=\frac{1}{a^{(3,0)}_{\DD}},
\label{eq:flipLO}
\end{equation}
which is RG invariant by Eq.~\eqref{eq:flipinv} and basis independent, since both truncations of Sec.~\ref{sec:LOrel} share the flip relations. Equation~\eqref{eq:flipLO} is the prediction of this section. Its first correction follows from the flip-odd part $\tfrac12[S(S+1)-I(I+1)]\,g(\Mop)$ of Sec.~\ref{sec:LOrel}. In the single-channel sectors $(0,3)$ and $(3,0)$ one has $\Mop=-\tfrac{15}4$ and $C^{(0,3)}_{\DD\DD}-C^{(3,0)}_{\DD\DD}=-12\,g(-\tfrac{15}4)$. In the $\NN$ channel $C^{(0,1)}_{\NN\NN}-C^{(1,0)}_{\NN\NN}=-2[\cos^2\theta_{\mathrm A}\,g(\lambda_+)+\sin^2\theta_{\mathrm A}\,g(\lambda_-)]$, with $\lambda_\pm=\tfrac54\pm2\sqrt5$. The ratio of the two differences,
\begin{equation}
\zeta\equiv\frac{6\,g(-\tfrac{15}4)}{\cos^2\theta_{\mathrm A}\,g(\lambda_+)+\sin^2\theta_{\mathrm A}\,g(\lambda_-)},
\label{eq:zeta}
\end{equation}
is not fixed by the $1/\Nc$ expansion. The minimal scheme of Eq.~\eqref{eq:bil} takes $g$ constant and gives $\zeta=6$. The sign of $\zeta$ is positive whenever $g$ varies slowly between $-\tfrac{15}4$ and $\lambda_-=-3.22$, which carries $87\%$ of the $\NN$ weight, so that the sign is a weaker assumption than the factor 6. We use the minimal scheme below and regard both the factor 6 and the sign as model statements. In the minimal scheme the first correction is Eq.~\eqref{eq:ratios} imposed on the couplings at $\LH$,
\begin{equation}
\begin{aligned}
C^{(0,3)}_{\DD\DD}(\LH)&=C^{(3,0)}_{\DD\DD}(\LH)+6\,\delta_{\NN}(\LH),\\
\delta_{\NN}&\equiv C^{(0,1)}_{\NN\NN}-C^{(1,0)}_{\NN\NN},
\end{aligned}
\label{eq:flipC}
\end{equation}
with every coupling given by Eq.~\eqref{eq:CLambda} in its channel. Since the two channels have the same loop, the relation is cutoff independent when written on the threshold matrices,
\begin{equation}
K^{(0,3)}_{\DD}-K^{(3,0)}_{\DD}=-\frac{6\,\delta_{\NN}(\LH)}{C^{(3,0)}_{\DD\DD}(\LH)\,C^{(0,3)}_{\DD\DD}(\LH)},
\label{eq:dKfinite}
\end{equation}
but it depends on $\LH$ through the couplings on the right-hand side, because Eq.~\eqref{eq:flipC} is imposed on couplings and not on threshold matrices. The size of the flip-odd term relative to a natural $\DD$ coupling, $1/a^{(3,0)}=0$, is
\begin{equation}
\frac{6\,\delta_{\NN}(\LH)}{C^{(3,0)}_{\DD\DD}(\LH)}=-0.76,\ -0.38,\ -0.19
\label{eq:flipsize}
\end{equation}
at $\LH=770$, $1500$ and $3000$~MeV, scaling as $1/\LH$, because $\delta_{\NN}\propto1/\LH^2$ and $C\propto1/\LH$. The flip-odd term is thus not a small correction at the vector-meson scale, a direct consequence of the factor 6 carried by the $\Delta$ quantum numbers, and at $\LH=770$~MeV Eq.~\eqref{eq:dKfinite} has a pole at $1/a^{(3,0)}=-152$~MeV, where $C^{(0,3)}_{\DD\DD}(\LH)$ changes sign. Tab.~\ref{tab:flipband} gives the offset $1/a^{(0,3)}-1/a^{(3,0)}$ for $1/a^{(3,0)}=0$ and $100$~MeV as $\LH$ is varied, which is the band within which the minimal-scheme correction is defined. A scale-independent form is obtained for $\LH\to\infty$, where every diagonal coupling approaches $-\pi^2/(\mu_i\LH)$ and $\delta K=-C^{-1}\delta C\,C^{-1}$ becomes exact, and reads on the threshold matrices
\begin{equation}
\begin{aligned}
K^{(S,I)}_{\alpha\alpha}-K^{(I,S)}_{\alpha\alpha}&=-\frac{S(S+1)-I(I+1)}{2}\Big(\frac{\mu_\alpha}{\mu_1}\Big)^2\,\delta K_{\NN},\\
\delta K_{\NN}&\equiv K^{(0,1)}_{\NN\NN}-K^{(1,0)}_{\NN\NN}=\frac{\mu_1}{2\pi}\Big(\frac1{a_s}-\frac1{a_t}\Big),
\end{aligned}
\label{eq:dK}
\end{equation}
with $\mu_\alpha$ the reduced mass of channel $\alpha$ and $\delta K_{\NN}=-3.34\times10^3$~MeV$^2$. The factors $-\tfrac12[S(S+1)-I(I+1)](\mu_\alpha/\mu_1)^2$ are $1.72$ for $\DD$ in the $\NN$ sectors, $2.58$ and $3.44$ for $\ND$ and $\DD$ in $(1,2)$--$(2,1)$, $10.3$ for $(0,3)$--$(3,0)$ and $5.17$ for $(2,3)$--$(3,2)$. For the single-channel pair, $K=(\mu_2/2\pi)(1/a)$ and
\begin{equation}
\frac{1}{a^{(0,3)}_{\DD}}-\frac{1}{a^{(3,0)}_{\DD}}=6\,\frac{\mu_2}{\mu_1}\Big(\frac1{a_s}-\frac1{a_t}\Big)=-352~\text{MeV}.
\label{eq:flipasym}
\end{equation}

\begin{table}[t]
\caption{The minimal-scheme offset $1/a^{(0,3)}_{\DD}-1/a^{(3,0)}_{\DD}$ (MeV) from Eq.~\eqref{eq:dKfinite} for two values of $1/a^{(3,0)}_{\DD}$, as a function of the scale $\LH$ at which Eq.~\eqref{eq:flipC} is imposed, and the limit of Eq.~\eqref{eq:flipasym}. The last column is the size of the flip-odd term relative to the natural coupling, Eq.~\eqref{eq:flipsize}.}
\label{tab:flipband}
\begin{ruledtabular}\footnotesize\setlength{\tabcolsep}{4pt}
\begin{tabular}{cccc}
$\LH$ (MeV) & $1/a^{(3,0)}=0$ & $1/a^{(3,0)}=100$ & $6\delta_{\NN}/C^{(3,0)}$\\
\hline
770 & $-1578$ & $-603$ & $-0.76$\\
1000 & $-876$ & $-512$ & $-0.58$\\
1500 & $-585$ & $-441$ & $-0.38$\\
3000 & $-440$ & $-390$ & $-0.19$\\
$\infty$ & $-352$ & $-352$ & $0$\\
\end{tabular}
\end{ruledtabular}
\end{table}

Equation~\eqref{eq:flipasym} is a relation among scale-independent quantities, which is the form in which a large-$\Nc$ statement can be confronted with scattering lengths, but three qualifications apply. First, its derivation assumes the minimal scheme, and the $1/\Nc$ expansion fixes neither the size of $\zeta$ in Eq.~\eqref{eq:zeta} nor, without an assumption on the smoothness of $g$, its sign. The sign of Eq.~\eqref{eq:flipasym}, that the $(0,3)$ channel is less attractive than the $(3,0)$ channel, holds for $\zeta>0$. Second, the approach to the limit $\LH\to\infty$ is slow, Tab.~\ref{tab:flipband}, and at the vector-meson scale the offset differs from Eq.~\eqref{eq:flipasym} by a factor of four. Third, in nature the $\Delta$ width is not negligible in these channels, and a deeply bound $(3,0)$ state would have a binding momentum above $m_\pi$, outside the pionless theory. What survives as a prediction of the $1/\Nc$ expansion is therefore the LO equality of Eq.~\eqref{eq:flipLO}, with a first correction whose natural size is set by $6(\mu_2/\mu_1)|1/a_s-1/a_t|\simeq350$~MeV in the minimal scheme and whose sign follows only with an assumption on the smoothness of $g(\Mop)$.

The natural test is lattice QCD at heavy pion masses, where the $\Delta$ is stable, with the two $\DD$ scattering lengths and the $\NN$ scattering lengths all taken at the same pion mass. In flavor-SU(3)-symmetric lattice QCD the two channels belong to different irreducible representations of $\mathbf{10}\otimes\mathbf{10}$. The $(0,3)$ channel, with antisymmetric spin and symmetric flavor, lies in the $\mathbf{28}$ together with the $\Omega\Omega$ $\pw{1}{S}{0}$ channel, and the $(3,0)$ channel, with symmetric spin and antisymmetric flavor, lies in the $\overline{\mathbf{10}}$. The $(3,0)$ channel has been computed by the HAL QCD Collaboration at $m_\pi=679$--$1018$~MeV, where it is quasi-bound by $25$--$40$~MeV~\cite{gongyo2020}, and the $\Omega\Omega$ $\pw{1}{S}{0}$ channel of the $\mathbf{28}$ near the physical point~\cite{gongyo2018}. An SU(3)-symmetric calculation of the $\mathbf{28}$ and $\overline{\mathbf{10}}$ channels would test the SU(3) version of Eq.~\eqref{eq:flipLO} with the methods already in use for both. The $\NN$ scattering lengths at the same pion masses, which enter the first correction, are themselves the subject of disagreement between lattice collaborations, which limits a test of the correction but not of the LO equality.

%% ================================================================
\section{Conclusions}\label{sec:summary}
%% ================================================================

We have studied the LO contact interactions among nucleons and $\Delta$ isobars under a Wilsonian RG flow with a sharp momentum cutoff, keeping the $\Delta$ as an explicit coupled channel, and we have asked which large-$\Nc$ relations among these couplings are invariant under the flow and what controls the scale dependence of the others. The two-channel flow with the $\DD$ threshold inside the cutoff range is solved exactly, with the full matrix of threshold amplitudes as the renormalization condition. The closed channels decouple with relative corrections of order $\Lambda^2/\kappa^2$, the matching correction between the $\Delta$-full and the nucleon-only theory is $0.2\%$ to $4\%$ of the bare inverse coupling between $100$ and $600$~MeV, and the data transmitted to lower cutoffs are the invariants $\eta_K$ and $\bar C_K$ of the threshold matrix, which coincide with the transition ratio and the Schur complement of the couplings only in the decoupled regime and not at the vector-meson scale. The complete LO S-wave contact sector has ten $(S,I)$ sectors and 18 couplings, of which the twelve of Ref.~\cite{sk2020} parametrize the six that enter $\NN$ scattering, and the $\ND$ channel does not couple to $\NN$ in any S-wave.

On the group-theoretical side, the $1/\Nc$ expansion of the Hartree Hamiltonian applied to the $N$--$\Delta$ system, with the symmetric SU(4) representation constructed for $\Nc=3$ to $13$, gives the general-$\Nc$ matrix elements of the leading operator including the $N\leftrightarrow\Delta$ transitions. A generalized time reversal removes every correction of relative order $1/\Nc$, the LO is a function of $G_1\!\cdot\!G_2$ with seven parameters and eleven relations among the 18 couplings at $\Nc=3$, two relations survive at relative order $1/\Nc^2$, and the exact-SU(4) two-parameter form used in the literature differs from the LO by an operator of relative order $1/\Nc^2$ whose matrix elements in the $\Delta$ channels are of order one.

Under the flow, the eight flip relations $C^{(S,I)}=C^{(I,S)}$ are exact invariants, because the two sectors of a flip pair have identical loops, also when the central part of one-pion exchange is iterated in the S-waves. They are the large-$\Nc$ input for the $\Delta$ channels that can be imposed at any scale. The eigenchannel relations of the $\NN$--$\DD$ sectors hold only at the scale at which they are imposed. Their running is set by the closed-channel momentum $\kDD=\sqrt{2\MD\dM}$, which is of order $\Nc^0$, and by a parameter of the boundary condition that the $\NN$ data do not fix, and it is of order $10^\circ$ to $30^\circ$ in the eigenchannel angle between $770$ and $140$~MeV. The running persists at large $\Nc$, where the Hartree counting enhances it by $\Nc^2$ relative to the splitting of the eigen-couplings, so that a consistent large-$\Nc$ counting maximizes the violation of these relations by the flow. It is removed only in the formal limit $\kDD\to0$, which is the limit taken in Ref.~\cite{rss2025}. Relations among closed channels only are invariant in the decoupled regime and violated at the $10\%$ level between the two cutoffs considered here.

Two further results are negative in character. The momentum-independent contact terms generated by $\rho$, $\omega$ and scalar exchange at order $q^0$ contain no $N\to\Delta$ transition coupling, but this is an artifact of the $q^0$ truncation at $q\sim m_V$, which retains the exchanges that are subleading in $1/\Nc$ and drops the pion and the tensor coupling of the $\rho$, and the flip-odd part of this boundary condition exceeds the one implied by the LO couplings by more than an order of magnitude. A quantitative saturation analysis of the $\Delta$ channels requires the full one-boson-exchange potential at $q\sim\LH$. The flip relation between the $(S,I)=(3,0)$ and $(0,3)$ channels of $\DD$ scattering predicts equal scattering lengths at LO, a basis-independent and RG-invariant statement. In the minimal scheme its first correction is proportional to the Wigner-symmetry breaking of the $\NN$ scattering lengths, with a natural size of $350$~MeV in $1/a$ and a sign that makes the $(0,3)$ channel less attractive, but the $1/\Nc$ expansion fixes neither the size nor, without a smoothness assumption on the flip-odd function, the sign at this order.

The results of this work are statements about the contact sector. Between the cutoffs considered here the one-pion exchange is a dynamical degree of freedom and not a small perturbation on the S-wave contact above a few hundred MeV, so that the inclusion of the pion and of the $\pi N\Delta$ coupling in a distorted-wave flow is the natural continuation of this work. It would generate the transition couplings of the $\Delta$ channels that the contact flow cannot, and a coupled-channel analysis along the large-$\Nc$ trajectory with fixed $\kDD$ would address the potential-versus-amplitude question with an explicit $\Delta$. Lattice QCD calculations of the $I=3$ and $S=3$ $\DD$ scattering lengths at heavy pion masses, where the $\Delta$ is stable, and of their SU(3) partners in the $\mathbf{28}$ and $\overline{\mathbf{10}}$, can test the flip relation obtained here.

\begin{acknowledgments}
D.F.V.~gratefully acknowledges the Department of Science and Technology--Science Education Institute (DOST--SEI) for scholarship support and the Khon Kaen Particle Physics and Cosmology Theory Group (KKPaCT) at Khon Kaen University, Thailand, for hosting the Research Enhancement Program (SANDWICH). J.B.M.~acknowledges the generous financial support from Mindanao State University - Iligan Institute of Technology. DS has received funding support from the Fundamental Fund of Khon Kaen University. Generative AI tools (Claude, Anthropic, and ChatGPT, OpenAI) were used to assist with the drafting and grammatical correction of the manuscript text and with numerical implementation. The authors directed this assistance, verified all scientific content and equations against their own calculations, and take full responsibility for the paper.
\end{acknowledgments}

%% ================================================================
\appendix
\section{Conventions and spin--isospin matrices}\label{app:conv}
%% ================================================================

Natural units $\hbar=c=1$ are used, with $\hbar c=197.327$~MeV\,fm. The kinematics is nonrelativistic in every channel $i$, with reduced mass $\mu_i$, threshold energy $E_i$ measured from the $\NN$ threshold, and orbital angular momentum $\ell_i$. The inputs are $\MN=938.9$~MeV, $\MD=1232$~MeV, $\dM=\MD-\MN=293.1$~MeV, $\mu_{\NN}=\MN/2=469.5$~MeV, $\mu_{\ND}=\MN\MD/(\MN+\MD)=532.8$~MeV and $\mu_{\DD}=\MD/2=616.0$~MeV. The $\NN$ scattering lengths are $a_s=-23.74$~fm and $a_t=5.42$~fm, so that $1/a_s=-8.31$~MeV, $1/a_t=36.4$~MeV and $1/a_s-1/a_t=-44.7$~MeV, the physical effective ranges quoted in Sec.~\ref{sec:single} are $r_s=2.75$~fm and $r_t=1.75$~fm, and the pion parameters of Sec.~\ref{sec:window} are $m_\pi=138$~MeV, $g_A=1.27$ and $f_\pi=92.4$~MeV. In the two-channel $\NN$--$\DD$ sectors we write $\mu_1=\mu_{\NN}$ and $\mu_2=\mu_{\DD}$. The closed-channel momenta at the $\NN$ threshold are
\begin{equation}
\begin{aligned}
\kND&\equiv\sqrt{2\mu_{\ND}\dM}=558.9~\text{MeV},\\
\kDD&\equiv\sqrt{2\mu_{\DD}\cdot2\dM}=\sqrt{2\MD\dM}=849.8~\text{MeV}.
\end{aligned}
\label{eq:kappas}
\end{equation}
Partial waves are labeled ${}^{2S+1}L_J$ with total isospin $I$. Two-baryon spin and isospin states are coupled with Condon--Shortley phases, and the first particle label refers to the first baryon in the channel name. For identical baryons ($\NN$, $\DD$) the Pauli principle requires $L+S+I$ odd, and the coupled states $|\NN;S,I\rangle$ and $|\DD;S,I\rangle$ with $S+I$ odd are antisymmetric as they stand. The $\ND$ channel carries no Pauli constraint. The coupled state $|\ND;S,I\rangle$ with the nucleon as particle~1 is not an eigenstate of the exchange $P_{12}$, and the single $\ND$ channel used in this work is the normalized combination $|\ND\rangle_A=(|\ND;S,I\rangle-P_{12}|\ND;S,I\rangle)/\sqrt2$ with the exchange symmetry of the $\DD$ state, $P_{12}|\DD;S,I\rangle=-|\DD;S,I\rangle$ for $S+I$ odd. For a contact operator $O$ symmetric under $1\leftrightarrow2$ this gives $\langle\DD;S,I|O|\ND\rangle_A=\tfrac1{\sqrt2}[\langle\DD|O|\ND\rangle-\langle P_{12}\DD|O|\ND\rangle]=\sqrt2\,\langle\DD;S,I|O|\ND;S,I\rangle$, which is the factor $\sqrt2$ of Tab.~\ref{tab:dict}. The orthogonal combination does not couple to $\DD$ at LO. In the four-channel formulation of Ref.~\cite{sk2020}, in which $\ND$ and $\Delta N$ are kept as separate channels with equal couplings to $\DD$, the same factor arises from the coherent sum of the two.

The scattering amplitude in an open channel is $\mathcal A=(2\pi/\mu_i)(k\cot\delta-ik)^{-1}$ with $k\cot\delta=-1/a+\tfrac12rk^2+v_2k^4+\dots$. The tree amplitude is $\mathcal A=-V$, so that a positive potential is repulsive, and the Lippmann--Schwinger equation in a sector of coupled channels reads
\begin{equation}
\begin{aligned}
\mathcal A(E)&=-\big[V^{-1}-B(E;\Lambda)\big]^{-1},\qquad B_{ij}=\delta_{ij}B_i,\\
B_i(E;\Lambda)&=\int^{\Lambda}\!\frac{\dd^3q}{(2\pi)^3}\,\frac{1}{E-E_i-q^2/(2\mu_i)+i\varepsilon}.
\end{aligned}
\label{eq:LS}
\end{equation}
Plane waves are normalized to $(2\pi)^3\delta^3(\bm p-\bm p')$. The regulator is a sharp cutoff on the loop momentum, $\theta(\Lambda-q)$, applied in every channel. At LO the Wilsonian flow with this regulator is equivalent to subtractive renormalization, which is not the case beyond LO~\cite{egm2017}. App.~\ref{app:int} gives the integrals, and $B$ denotes the loop integral throughout, the letter $I$ being reserved for isospin. The momentum-space potentials of Ref.~\cite{sk2020} are defined with the opposite overall sign, such that a positive potential is attractive (their $K$ matrix gives $\delta=\arctan(\rho_{\rm ps}K)$ with a positive phase-space factor $\rho_{\rm ps}$), and their regulator is a local Gaussian on the meson propagators together with a nonlocal one on the contact terms. This is the origin of the sign dictionary of Eq.~\eqref{eq:signdict}.

The spin matrices of Ref.~\cite{sk2020} are, in the basis of decreasing magnetic quantum number,
\begin{equation}
\begin{aligned}
S^1&=\frac{1}{\sqrt6}\begin{pmatrix}-\sqrt3&0&1&0\\0&-1&0&\sqrt3\end{pmatrix},\\
S^2&=\frac{-i}{\sqrt6}\begin{pmatrix}\sqrt3&0&1&0\\0&1&0&\sqrt3\end{pmatrix},\\
S^3&=\frac{1}{\sqrt6}\begin{pmatrix}0&2&0&0\\0&0&2&0\end{pmatrix},
\end{aligned}
\label{eq:Smat}
\end{equation}
for the $\Delta\to N$ transition, with $S^{i\dagger}$ for $N\to\Delta$, and
\begin{equation}
\begin{aligned}
\Sigma^1&=\begin{pmatrix}0&\sqrt3&0&0\\ \sqrt3&0&2&0\\0&2&0&\sqrt3\\0&0&\sqrt3&0\end{pmatrix},\\
\Sigma^2&=\begin{pmatrix}0&-\sqrt3i&0&0\\ \sqrt3i&0&-2i&0\\0&2i&0&-\sqrt3i\\0&0&\sqrt3i&0\end{pmatrix},\\
\Sigma^3&=\mathrm{diag}(3,1,-1,-3),
\end{aligned}
\label{eq:Sigmat}
\end{equation}
for the $\Delta$, together with the Pauli matrices $\sigma^i$ for the nucleon. The tensors are
\begin{equation}
\begin{aligned}
S^{ij\dagger}&=-\frac{1}{\sqrt6}\big(S^{i\dagger}\sigma^j+S^{j\dagger}\sigma^i\big),\\
\Sigma^{ij}&=\frac18\big(\Sigma^i\Sigma^j+\Sigma^j\Sigma^i-10\delta^{ij}\big),\\
\Sigma^{ijk}&=\frac{1}{36\sqrt3}\Big[5\sum_{\rm perm}\Sigma^i\Sigma^j\Sigma^k\\
&\qquad-82\big(\Sigma^i\delta^{jk}+\Sigma^j\delta^{ik}+\Sigma^k\delta^{ij}\big)\Big],
\end{aligned}
\label{eq:tensors}
\end{equation}
where the sum runs over the six permutations of $(i,j,k)$. The isospin matrices $\tau^a$, $T^{a\dagger}$ and $\Theta^a$ are given by the same matrices in isospin space. The products satisfy $S^iS^{j\dagger}=\tfrac13(2\delta^{ij}-i\epsilon^{ijk}\sigma^k)$, $S^{i\dagger}S^j=\tfrac13\delta^{ij}-\tfrac13\Sigma^{ij}+\tfrac16i\epsilon^{ijk}\Sigma^k$ and $\Sigma^i\Sigma^j=5\delta^{ij}+4\Sigma^{ij}+i\epsilon^{ijk}\Sigma^k$, and $\Sigma^{ij}$ and $\Sigma^{ijk}$ are symmetric and traceless, which we have verified numerically. The relation to the large-$\Nc$ generators at $\Nc=3$ is given in App.~\ref{app:su4}.

%% ================================================================
\section{Cutoff integrals}\label{app:int}
%% ================================================================

With $k^2=2\mu(E-E_i)$ and a sharp cutoff, the loop integral of a channel with orbital angular momentum $\ell$ and separable form factor $q^\ell$ is
\begin{equation}
B^{(\ell)}(E;\Lambda)=\frac{\mu}{\pi^2}\int_0^\Lambda\!\dd q\,\frac{q^{2+2\ell}}{k^2-q^2+i\varepsilon}.
\label{eq:Il}
\end{equation}
For $\ell=0$ and $k^2>0$, $k<\Lambda$, the principal value of $\int_0^\Lambda\dd q/(k^2-q^2)$ is $\tfrac1{2k}\ln\tfrac{\Lambda+k}{\Lambda-k}$ and the delta function gives $-i\pi/(2k)$, which yields Eq.~\eqref{eq:Iopen}. For $k^2=-\kappa^2<0$, $\int_0^\Lambda\dd q/(\kappa^2+q^2)=\tfrac1\kappa\arctan\tfrac\Lambda\kappa$, which yields Eq.~\eqref{eq:Iclosed}. The continuation of Eq.~\eqref{eq:Iopen} with $k=i\kappa$ gives $\tfrac k2\ln\tfrac{\Lambda+k}{\Lambda-k}\to-\kappa\arctan\tfrac\kappa\Lambda$ and $-i\mu k/(2\pi)\to\mu\kappa/(2\pi)$, and with $\arctan(\kappa/\Lambda)=\pi/2-\arctan(\Lambda/\kappa)$ the two terms combine into Eq.~\eqref{eq:Iclosed}. For $\ell=2$ one uses $q^6/(k^2-q^2)=-q^4-k^2q^2-k^4+k^6/(k^2-q^2)$ and obtains
\begin{equation}
\begin{aligned}
B^{(2)}(E;\Lambda)&=-\frac{\mu}{\pi^2}\Big[\frac{\Lambda^5}{5}+\frac{k^2\Lambda^3}{3}+k^4\Lambda-\frac{k^5}{2}\ln\frac{\Lambda+k}{\Lambda-k}\Big]\\
&\quad-\frac{i\mu k^5}{2\pi}.
\end{aligned}
\label{eq:ID}
\end{equation}
In all cases $\partial_\Lambda B^{(\ell)}=-(\mu/\pi^2)\Lambda^{2+2\ell}/(\Lambda^2-k^2)$, which is the only property of the regulator used in the flow equations. For a smooth regulator $f(q/\Lambda)$ the derivative is replaced by the convolution of the same integrand with $-\partial_\Lambda f$, and the decoupling statements of Sec.~\ref{sec:decoupling} hold with regulator-dependent coefficients.

%% ================================================================
\section{Wilsonian form of the flow, fixed points, and the $2^+$ system}\label{app:wilson}
%% ================================================================

From $\partial_\Lambda V^{-1}=\partial_\Lambda B$ one has $\partial_\Lambda V=V\,D\,V$ with $D$ diagonal, $D_{ii}=(\mu_i/\pi^2)\Lambda^{2+2\ell_i}/(\Lambda^2-k_i^2)$. With the rescaling $\Vh_{ij}=(\Lambda^{1+\ell_i+\ell_j}/\pi^2)\sqrt{\mu_i\mu_j}\,V_{ij}$, which for S-waves is $\Vh=(\Lambda/\pi^2)M^{1/2}VM^{1/2}$ with $M=\mathrm{diag}(\mu_i)$, together with $\kh_i=k_i/\Lambda$ and $\kaph_i=\kappa_i(0)/\Lambda$,
\begin{equation}
\Lambda\partial_\Lambda\Vh_{ij}\big|_k=(1+\ell_i+\ell_j)\Vh_{ij}+\sum_m\Vh_{im}\frac{1}{1-\kh_m^2}\Vh_{mj},
\end{equation}
since $\Lambda^{1+\ell_i+\ell_j}\Lambda^{3+2\ell_m}/\Lambda^{2+\ell_i+\ell_j+2\ell_m}=\Lambda^2$. Changing variables from $(k_1,\kappa_i,\Lambda)$ to $(\kh_1,\kaph_i,\Lambda)$ at fixed $\kh$ and $\kaph$ adds the Euler term, and the Wilsonian equation for channels of arbitrary partial waves reads
\begin{equation}
\Lambda\frac{\partial\Vh}{\partial\Lambda}=\Vh+\mathcal L\Vh+\Vh\mathcal L+\Eu\Vh+\Vh\Dh\Vh,\qquad
\mathcal L=\mathrm{diag}(\ell_i),
\label{eq:birseL}
\end{equation}
with $\Dh_{ii}=1/(1-\kh_i^2)$ and $\Eu=\kh_1\partial_{\kh_1}+\sum_i\kaph_i\partial_{\kaph_i}$. For two S-wave channels, $\ell_i=0$, it reads $\Lambda\partial_\Lambda\Vh=\Vh+\Eu\Vh+\Vh\Dh\Vh$ with $\kh_2^2=(\mu_2/\mu_1)\kh_1^2-\kaph^2$, which is the form used in Sec.~\ref{sec:fixedpoints}. For $W=\Vh^{-1}$ the equation is linear, $\Lambda\partial_\Lambda W=-W-\mathcal LW-W\mathcal L+\Eu W-\Dh$, and fixed points satisfy $(1-\Eu)W+\mathcal LW+W\mathcal L=-\Dh$. Since $(1+2\ell_i-\Eu)(\kh_i^2)^n=(1+2\ell_i-2n)(\kh_i^2)^n$, the diagonal solutions are $W_{ii}=-\varphi_{\ell_i}(\kh_i^2)$ with
\begin{equation}
\begin{aligned}
\varphi_\ell(x)&=\sum_{n\ge0}\frac{x^n}{2\ell+1-2n},\qquad \varphi_0(x)=1-\sqrt x\,\mathrm{artanh}\sqrt x,\\
\varphi_2(x)&=\frac15+\frac x3+x^2-\frac{x^{5/2}}{2}\ln\frac{1+\sqrt x}{1-\sqrt x}.
\end{aligned}
\label{eq:fl}
\end{equation}
The off-diagonal equations are homogeneous, $(1+\ell_i+\ell_j-\Eu)W_{ij}=0$, so that $W_{ij}$ is homogeneous of degree $1+\ell_i+\ell_j$ in $(\kh,\kaph)$. Analyticity in $E$ and in the mass splittings requires even degree, so that $W_{ij}=0$ whenever $\ell_i+\ell_j$ is even, and every analytic fixed point with invertible $\Vh_*$ is diagonal in the partial-wave label. The analysis in terms of $W=\Vh^{-1}$ does not cover rank-deficient fixed points, for which $\Vh_*$ is not invertible, and a rank-deficient point that mixes different partial waves is not excluded by it. For S-waves the rank-one solutions of Eq.~\eqref{eq:rank1} follow by inserting $\Vh=v\,nn^{\Tp}$ with constant $n$, which gives $1/v=-\sum_in_i^2\varphi_0(\kh_i^2)$. Linearizing around a fixed point, $\Vh=\Vh_*+\varepsilon$, with $\Vh_*\to-\mathrm{diag}[p_i(2\ell_i+1)]$ at $\kh,\kaph\to0$ and $p_i=1$ for a channel at its nontrivial point and $0$ otherwise, a component of degree $d$ obeys $\Lambda\partial_\Lambda\varepsilon_{ij}=\nu_{ij}\varepsilon_{ij}$ with
\begin{equation}
\nu_{ij}=d+1+\ell_i+\ell_j-(2\ell_i+1)p_i-(2\ell_j+1)p_j,
\label{eq:nuL}
\end{equation}
which for $\ell=0$ gives $\nu=d-1$ at the rank-two point, $d+1$ at the trivial point, and $d+1-p_a-p_b$ in general, so that around the $\NN$-aligned rank-one point ($p_1=1$, $p_2=0$) the $d=0$ couplings have $\nu=-1,0,+1$ and the $d=2$ perturbations, among them the explicit $\kaph^2$ dependence on the mass splitting, have $\nu=+1,+2,+3$, and the rank-two point reproduces the $\Lambda^{-2}$ scaling of the mixing coupling of Ref.~\cite{cgvk2004}. The dimensions at finite $\kh$ are the same, because $\Eu$ is diagonal on homogeneous functions and $\Dh-\bone$ mixes a given degree only with higher degrees. The fixed points were checked numerically: the residuals of the two-channel equation for the rank-one and rank-two points at four values of $(\kh_1,\kaph)$ are below $10^{-10}$, and those of Eq.~\eqref{eq:birseL} for the $2^+$ point below are below $4\times10^{-10}$.

The $J^P=2^+$, $I=1$ system contains the channels $1=\NN\,\pw{1}{D}{2}$ ($\ell_1=2$, open), $2=\ND\,\pw{5}{S}{2}$ ($\ell_2=0$, closed for $E<\dM$, with $\kappa_2(E)=\sqrt{2\mu_{\ND}(\dM-E)}$) and $3=\DD\,\pw{5}{S}{2}$ ($\ell_3=0$, closed, with $\kappa_3(E)=\sqrt{2\mu_{\DD}(2\dM-E)}$). The lowest-dimension contact operators, separable in the form factors $p^{\ell_i}$, give
\begin{equation}
V(p',p)=\begin{pmatrix}C_{11}p'^2p^2 & C_{12}p'^2 & C_{13}p'^2\\ C_{12}p^2 & C_{22} & C_{23}\\ C_{13}p^2 & C_{23} & C_{33}\end{pmatrix},
\label{eq:V3}
\end{equation}
with the D-wave tensor structure of the $\NN\to\ND$ and $\NN\to\DD$ transitions absorbed into the projected constants $C_{12}$ and $C_{13}$, which in this appendix denote the couplings of the three-channel system and not the $\NN$--$\DD$ couplings of the main text. The Lippmann--Schwinger equation is algebraic, $T=g(p')^{\Tp}[C^{-1}-B(E;\Lambda)]^{-1}g(p)$ with $B=\mathrm{diag}(B^{(2)}_1,B^{(0)}_2,B^{(0)}_3)$, and the flow is $\Lambda\,\dd C/\dd\Lambda=C\mathcal DC$ with $\mathcal D=\mathrm{diag}[(\mu_1/\pi^2)\Lambda^7/(\Lambda^2-k_1^2),(\mu_2/\pi^2)\Lambda^3/(\Lambda^2+\kappa_2^2),(\mu_3/\pi^2)\Lambda^3/(\Lambda^2+\kappa_3^2)]$. By the diagonality result above, the analytic fixed points are products of a D-wave point for channel~1, trivial or $W_{11}=-\varphi_2$, and a point of the two-channel S-wave family of Sec.~\ref{sec:fixedpoints} for channels~2 and~3. In the physical regime $\Lambda<\kDD$ the function $\varphi_0(\kh_2^2)$ with $\kh_2^2<0$ continues to $1+\sqrt y\arctan\sqrt y$ with $y=-\kh_2^2$, whose large-$y$ behavior $\tfrac\pi2\kaph+1/(3\kaph^2)$ contains the degree-one piece $\mu_2\kappa/2\pi$ of Eq.~\eqref{eq:Ilimits}, which belongs to the amplitude and not to the potential, and the decoupling term $\Lambda^2/(3\kDD^2)$. At the physical point for $E\ll\dM$ the $\NN$ D wave is natural ($p_1=0$) and the closed channels are at their trivial points, so that $\nu(C_{11})=5$, $\nu(C_{12})=\nu(C_{13})=3$ and $\nu(C_{22},C_{23},C_{33})=1$: every coupling of the system is irrelevant and the transition couplings are irrelevant by three orders, which is the counting statement that a two-derivative transition is of order $Q^2$ relative to an S-wave contact. Equation~\eqref{eq:nuL} also covers a near-threshold $\ND$ state, $p_2=1$ with $\nu(C_{12})=2$ and $\nu(C_{22})=-1$. The size of the $\ND$ effect in $\pw{1}{D}{2}$ follows by eliminating channel~2, $\delta C_{11}=C_{12}^2B^{(0)}_2(E)$ with the closed loop of Eq.~\eqref{eq:Iclosed}: for a transition coupling of natural size, $|C_{12}|\sim2\pi/(\sqrt{\mu_1\mu_{\ND}}\Lambda^3)$, this is $5\%$ of a natural $C_{11}\sim2\pi/(\mu_1\Lambda^5)$ at $\Lambda=300$~MeV and $15\%$ at $600$~MeV. The vanishing of the mixing coupling at every fixed point of a two-partial-wave Wilsonian flow was found for the pionless $\pw{3}{S}{1}$--$\pw{3}{D}{1}$ system by Harada and Kubo~\cite{hk2006}, and the result above is its generalization to arbitrary $\ell_i$ and to closed channels.

%% ================================================================
\section{SU(4) construction}\label{app:su4}
%% ================================================================

The totally symmetric representation of SU(4) with $\Nc$ boxes is realized by $\Nc$ bosons in the four single-quark states $|s,t\rangle$, with the one-body operators $X=\sum_{\alpha\beta}X_{\alpha\beta}a^\dagger_\alpha a_\beta$ for the single-quark matrices $\sigma^i/2\otimes\bone$, $\bone\otimes\tau^a/2$ and $\sigma^i\otimes\tau^a/4$. Its dimension is $\binom{\Nc+3}{3}$, equal to $20$ at $\Nc=3$. In the sector $J_z=I_z=j$ the operator $J^2$ has exactly one eigenvector with eigenvalue $j(j+1)$, the highest-weight state of the $J=I=j$ multiplet, and the remaining states follow by lowering with $J_-$ and $I_-$, which gives Condon--Shortley phases. The relative phase between the $j=\tfrac12$ and $j=\tfrac32$ multiplets is fixed so that $\langle\Delta^+\tfrac12|G^{33}|p\tfrac12\rangle<0$. Two-baryon channel states $|B_1B_2;S,I\rangle$ are built with Clebsch--Gordan coefficients on the highest weights, antisymmetrized under baryon exchange and normalized, and every two-body operator is applied as a sequence of one-body actions, so that intermediate states run over the full representation. This matters for products such as $\Mop^2$ at $\Nc\ge5$, where intermediate $j\ge\tfrac52$ states contribute.

Because the state is symmetric in spin--flavor, the spin and isospin exchange operators of any quark pair satisfy $P^\sigma_{qq'}P^\tau_{qq'}=1$, which gives $\bm\sigma_q\!\cdot\!\bm\sigma_{q'}=\bm\tau_q\!\cdot\!\bm\tau_{q'}$ for $q\ne q'$. From this pair identity follow the exact one-body relations
\begin{equation}
\{I^a,G^{ia}\}=\frac{\Nc+2}{2}J^i,\qquad \{J^i,G^{ia}\}=\frac{\Nc+2}{2}I^a,
\label{eq:IG}
\end{equation}
the reduction rule for two $G$'s with contracted spin indices,
\begin{equation}
\sum_iG^{ia}G^{ib}=\frac{\Nc(\Nc+4)}{16}\delta^{ab}-\frac14\delta^{ab}I^2+\frac14I^bI^a+\frac i2\epsilon^{abc}I^c,
\label{eq:GGreduce}
\end{equation}
whose symmetric traceless part is the Dashen--Jenkins--Manohar identity~\cite{djm1995}, and the Casimir relation $J^2+I^2+4G^2=3\Nc(\Nc+4)/4$. Within a multiplet $J=I=j$ the only rank-$(1,1)$ operator is $J^iI^a$, so that $G^{ia}=c_jJ^iI^a$, and the anticommutator with $I^a$ in Eq.~\eqref{eq:IG} gives $c_j=(\Nc+2)/[4j(j+1)]$. The transition strength follows from the Casimir, $\sum|\langle\Delta|G^{ia}|N\rangle|^2=3\Nc(\Nc+4)/16-3/8-(\Nc+2)^2/16=(\Nc-1)(\Nc+5)/8$. For a two-body product $X_1Y_2$ and the antisymmetrized $\ND$ state, $\langle A|X_1Y_2|A\rangle=\langle\ND|X_1Y_2|\ND\rangle-\langle\ND|X_1Y_2|\Delta N\rangle$, the direct term being $4c_{j_1}c_{j_2}\langle\Jop\rangle\langle\Iop\rangle$ for $\Mop$ with $\langle\Jop\rangle=\tfrac12[S(S+1)-j_1(j_1+1)-j_2(j_2+1)]$, and the exchange term requiring two $N\leftrightarrow\Delta$ transitions and scaling as $(\Nc-1)(\Nc+5)$. The polynomials of Tab.~\ref{tab:closed} were fixed from these forms and the $\Nc=3$ and $5$ values and verified at $\Nc=7,9,11,13$ to $10^{-14}$. At $\Nc=3$ the generators are expressed through the matrices of App.~\ref{app:conv} as $G^{ia}=\tfrac5{12}\sigma^i\tau^a$ on the nucleon, $G^{ia}=\tfrac1{12}\Sigma^i\Theta^a$ on the $\Delta$ and $G^{ia}=-\tfrac1{\sqrt2}S^{i\dagger}T^{a\dagger}$ for $N\to\Delta$, which reproduces every entry of Tab.~\ref{tab:closed} from Tables~\ref{tab:dict} and \ref{tab:factors}, with the $\ND$ diagonal elements including the exchange term. The transition element of the $(0,1)$ sector follows also by recoupling, with $\{\tfrac32\tfrac32 0;\tfrac12\tfrac12 1\}=-\sqrt2/4$ and $\{\tfrac32\tfrac32 1;\tfrac12\tfrac12 1\}=\sqrt{10}/12$, as $4\cdot\tfrac12\cdot(-\sqrt2)(-\sqrt{10}/3)=4\sqrt5/3$. The antisymmetric part of $20\otimes20$ decomposes into the SU(4) irreducible representations $140\oplus50$ with Casimirs $\tfrac{33}2$ and $\tfrac{21}2$, on which $\Bop=\tfrac34$ and $-\tfrac{21}4$. Matching the $(0,1)$ sector of the exact-SU(4) form to Eqs.~(49), (52) and (53) of Ref.~\cite{rss2025} gives $c_1=2\tilde a+\tilde b/2$ and $g=2\tilde b/9$ in their notation, and the remaining five of their relations are then reproduced without further input.

%% ================================================================
\section{The boundary-condition family}\label{app:family}
%% ================================================================

With $C^{-1}(\LH)=\alpha\bone+\beta\Mop$ and $K=\alpha\bone+\beta\Mop-B(\LH)$, the eigen-couplings are $1/v_\pm=\alpha+\beta\lambda_\pm$, and fixing their ratio $r_v=v_+/v_-$ gives $\alpha=c\beta$ with $c=(r_v\lambda_+-\lambda_-)/(1-r_v)$. Inserting $K$ into the scattering-length condition $K_{11}-K_{12}^2/K_{22}=s\equiv\mu_1/(2\pi a)$ gives, with $g_i=c+\Mop_{ii}$,
\begin{equation}
\begin{aligned}
&\big[(c+\lambda_+)(c+\lambda_-)\big]\beta^2-\big[g_1B_2(\LH)+g_2(B_1(\LH)+s)\big]\beta\\
&\quad+(B_1(\LH)+s)\,B_2(\LH)=0.
\end{aligned}
\label{eq:quadratic}
\end{equation}
Since $B_2(\LH)$ is small compared with $B_1(\LH)$ in the decoupled regime, one root is of order $B_2$ and corresponds to $K_{22}\to0$, a $\DD$ channel at a pole, and is discarded. The other is $\beta=g_2(B_1+s)/[(c+\lambda_+)(c+\lambda_-)]\,[1+\mathcal O(B_2/B_1)]$, which is Eq.~\eqref{eq:betahat}. The exact root is used in all numerical results. With $n_+=(\cos\theta_{\mathrm A},\sin\theta_{\mathrm A})$, $\cos\theta_{\mathrm A}\sin\theta_{\mathrm A}=\tfrac13$ and $\cos^2\theta_{\mathrm A}=(3-\sqrt5)/6$, the transition ratio at $\LH$ is
\begin{equation}
\eta(\LH)=\frac{(v_+-v_-)\cos\theta_{\mathrm A}\sin\theta_{\mathrm A}}{v_+\cos^2\theta_{\mathrm A}+v_-\sin^2\theta_{\mathrm A}}
=\frac{2(r_v-1)}{r_v(3-\sqrt5)+3+\sqrt5},
\end{equation}
which is Eq.~\eqref{eq:rho0}. For the closed $(1,2)$ sector the analogous boundary condition $C^{-1}(\LH)=\alpha'\bone+\beta'\Mop_{(1,2)}$ has no datum to fix $\alpha'$, and Fig.~\ref{fig:exact}(b) uses $\alpha'=-\mu_{\ND}\LH/\pi^2$, a coupling of natural size, with $\beta'=\bh'\mu_{\ND}\LH/\pi^2$.

%% ================================================================
\section{Low-momentum reduction of one-boson exchange between $N$ and $\Delta$ states}\label{app:obe}
%% ================================================================

We write the meson--baryon couplings in the form used in one-boson-exchange models, with a scalar $\sigma$ coupled as $g_\sigma\bar\psi\psi\sigma$, the $\omega$ coupled to the baryon-number current as $g_\omega\bar\psi\gamma^\mu\psi\omega_\mu$, and the $\rho$ coupled to the isospin current as $g_\rho\bar\psi\gamma^\mu I^a\psi\rho^a_\mu$ plus the tensor term $(f_\rho/2M)\bar\psi\sigma^{\mu\nu}I^a\psi\,\partial_\mu\rho^a_\nu$ with $\kappa_\rho=f_\rho/g_\rho$, for both the nucleon and the $\Delta$, with the $\Delta$ as a Rarita--Schwinger field. The couplings to the $\Delta$ are taken equal to those of the nucleon, which is the SU(6) relation and, for the $\omega$ and the vector coupling of the $\rho$, the statement that these mesons couple to the conserved baryon-number and isospin currents at LO in $1/\Nc$. In the nonrelativistic reduction to LO in $q/M$ the scalar and the time components of the vector exchanges are spin independent, and the exchange potentials between any two baryons of the multiplet are
\begin{equation}
V^{(0)}(q)=-\frac{g_\sigma^2}{q^2+m_\sigma^2}+\frac{g_\omega^2}{q^2+m_\omega^2}+\frac{g_\rho^2}{q^2+m_\rho^2}\,4I_1\!\cdot\!I_2,
\end{equation}
where $4I_1\!\cdot\!I_2$ has the eigenvalues $2[I(I+1)-i_1(i_1+1)-i_2(i_2+1)]$, namely $-3,+1$ for $\NN$ ($I=0,1$), $-5,+3$ for $\ND$ ($I=1,2$) and $-15,-11,-3,+9$ for $\DD$ ($I=0,\dots,3$), in agreement with Tab.~\ref{tab:factors}. Expanding the propagators for $q\ll m$ gives the $q^0$ couplings of Eq.~\eqref{eq:sat}. The spin-dependent terms of vector exchange arise at order $q^2/M^2$: the spin--orbit terms $i(\bm S_1+\bm S_2)\!\cdot\!(\bm q\times\bm k)$ and, from the magnetic couplings, $(1+\kappa)^2(\bm S_1\times\bm q)\!\cdot\!(\bm S_2\times\bm q)/(4M^2)$, with the appropriate spin operators of App.~\ref{app:conv} for each baryon pair. They are of order $q^2$ and map onto the $C^{(2)}$ structures of Eq.~\eqref{eq:ctNLO} and its $\Delta$ analogues. No $q^0$ structure connects the nucleon to the $\Delta$: the scalar and the charge-type vector vertices are proportional to $\bone$ and $I^a$, which are diagonal in the multiplet, whereas the magnetic $\rho N\Delta$ vertex and the $\pi N\Delta$ vertex are proportional to $S^{i\dagger}T^{a\dagger}q^i$ and vanish at $q=0$. This is the group-theoretical content of Eq.~\eqref{eq:notr}. Numerical values of the couplings, including $\kappa_\rho$, are not used in this work and have to be taken from a stated one-boson-exchange model when the boundary condition is evaluated.

\ifrevtex\bibliographystyle{apsrev4-2}\else\bibliographystyle{unsrt}\fi
\bibliography{refs}

\end{document}